# Unit Vectors for Similar Oblate Spheroidal Coordinates and Vector Transformation

Pavel Strunz

*Nuclear Physics Institute of the Czech Academy of Sciences, 25068 Řež, Czech Republic, strunz@ujf.cas.cz*

**ABSTRACT**

The unit vectors transformation between the Cartesian and the novel Similar Oblate Spheroidal coordinates, and vice versa, is derived. It can help to transform vector fields between these two types of orthogonal coordinates which can advantageously simplify solutions of problems exhibiting oblate spheroidal geometry. Several examples demonstrate the use of the derived relations. Generalized sine and cosine applicable in Similar Oblate Spheroidal coordinate system are introduced.

## 1   Introduction

Recently finalized analytical solution of the Similar Oblate Spheroidal (SOS) orthogonal coordinate system [1], [2] can be a powerfull tool for a description of physical processes inside or in the vicinity of the bodies with geometry of an oblate spheroid. Such bodies range (but are not limited to) from planets with a small oblateness (like the Earth with the semi-axes ratio ≈1:300), through elliptical galaxies up to significantly flattened objects like disk galaxies.

In the field of atmospheric physics, it was stated [1] that SOS coordinates could be of help for better modeling of geopotential surfaces, allowing for a better description of the spatial variation of apparent gravity. Further, similar oblate spheroids are frequently used for modelling of iso-density levels inside galaxies [3], [4]. Therefore the SOS coordinates can be helpful also in this field. It would be of advantage to have a possibility to transform between vectors (e.g. force vector, fluid velocity vector field) expressed in the SOS and in the Cartesian coordinates. Therefore, there exist a clear physical motivaton for the derivation of the formulas enabling such transformation.

Similar Oblate Spheroidal coordinates are distinct from the well known Confocal Oblate Spheroidal coordinates, which do not possess similarity between the individual spheroids within the family of spheroidal coordinate surfaces.

Although coordinate transformation from the SOS coordinates to the Cartesian system as well as metric scale factors and Jacobian determinant were already reported [2], a description of the transformation of a vector field **A** between the SOS system and the Cartesian coordinates is still missing.

The transformation of a vector field **A** between two orthogonal coordinate systems requires unit vectors to be determined. In case of the vector transformation from the Similar Oblate Spheroidal orthogonal coordinate system (coordinates denoted $R$, $v$, $λ$) to the Cartesian coordinate system (coordinates $x_{3D}$, $y_{3D}$, $z_{3D}$), the expressions for the unit vectors in the SOS system $\widehat{R}$, $\widehat{v}$, $\widehat{\lambda}$ has to be found.

The vector field **A** can be then written in terms of the unit vectors as

$$\mathbf{A} = A_x\hat{\mathbf{x}} + A_y\hat{\mathbf{y}} + A_z\hat{\mathbf{z}} = A_R\widehat{R} + A_v\widehat{v} + A_\lambda\widehat{\lambda} \quad , \tag{1.1}$$

where $\hat{\mathbf{x}}$, $\hat{\mathbf{y}}$, $\hat{\mathbf{z}}$ are the unit vectors in Cartesian coordinates. $A_x, A_y, A_z$ are the components of the vector **A** in the Cartesian coordinates, while $A_R$, $A_v$, $A_\lambda$ are the components of the vector **A** in the SOS coordinates. When unit vectors $\hat{\mathbf{x}}$, $\hat{\mathbf{y}}$, $\hat{\mathbf{z}}$ are selected to be equal to the basis vectors of the Cartesian coordinate system, as is done in this article, then the above relation represents way of vector **A** transformation from the SOS coordinates to the Cartesian coordinates.

The following text deals with the derivation of the unit vectors $\widehat{R}$, $\widehat{v}$, $\widehat{\lambda}$ of the SOS coordinate system, which then enables transformation of vectors and vector fields from the SOS coordinates to the Cartesian coordinates. The inverse transformation is dealt with as well: as both coordinate systems are orthogonal, the derivation of the inverse transformation relations does not pose further tedious work.

The organization of the text is as follows: First, the relevant formulas for the SOS coordinates and their transformations to the Cartesian coordinates [2] are summarized (chapter 2), then also the relevant combinatorial identities are reminded (chapter 3). The main parts of the article are chapters 4 and 5, in which the unit vectors are determined in two distinct regions of space. A matrix form of the unit vectors transformation is reported in chapter 6. Finally, several examples of the use of the unit vector transformation is shown (chapter 7).

## 2 SOS coordinates and their derivatives, metric scale factors and Jacobian

### 2.1 SOS coordinates

For SOS coordinates [1], the basic coordinate surfaces of the $R$ coordinate are similar oblate spheroids. They are generated by rotating similar ellipses, given in Cartesian coordinates by the formula

$$x^2 + (1 + \mu)\, z^2 = R^2 \, , \tag{2.1}$$

around the minor axis. The $R$ coordinate value is equal to the major semi-axis length (i.e. to the equatorial radius) of the particular spheroid from the family. The parameter $\mu$ characterizes the whole family of similar ellipses, as well as the similar oblate spheroids generated by rotating the ellipses, having the same eccentricity $e$:

$$e = \sqrt{\frac{\mu}{1+\mu}} \, , \quad . \tag{2.2}$$

The parameter $\mu>0$ for oblate spheroids, and the minor and major semi-axes of each member of the spheroid family have the ratio $(1+\mu)^{-1/2}$. As a limit (when $\mu = 0$) sphere is determined by Eq. (2.1) with corresponding spherical coordinate surface of the $R$ coordinate.

A special – called "reference" in what follows – spheroid is introduced with the equatorial radius $R_0$. Usually, this "reference spheroid" of the SOS coordinate system with the equatorial radius $R_0$ would coincide with the surface of the body under investigation for which the SOS coordinate system is applied.

The second set of coordinate surfaces, orthogonal to the similar oblate spheroids defined above, are power curves rotated around the minor axis of the above spheroids [1], [2]. The labeling, i.e. the coordinate *v*, corresponding to these surfaces is equivalent to the so called parametric latitude [2]. The coordinate *v* is also equivalent to the parameter used for standard parametric equation of ellipse with a special $R_0$ major semi-axis, i.e. $x = R_0 \cos v$ and $z = \frac{R_0}{\sqrt{1+\mu}} \sin v$. This reference ellipse generates the reference spheroid with the equatorial radius $R_0$ mentioned above. An example of the SOS coordinatre system section with *x-z* plane is displayed in Fig. 1.

Finally, the third set of coordinate surfaces, orthogonal to the previous two, are then semi-infinite planes containing the rotation axis. The associated coordinate is the longitude angle λ, which is the same as its equivalent coordinate in the spherical coordinate system.

A key role in the derivation of the SOS coordinate transformations plays the dimensionless parameter *W* defined as

$$W = \left(\frac{R}{R_0}\right)^\mu \frac{\sin v}{\cos^{1+\mu} v} \quad , \tag{2.3}$$

which will be used frequently later. In what follows, the calculation is restricted to the parametric latitude *v* positioned only in the first quadrant of the *x-z* plane of the Cartesian coordinate system, thus the generating ellipse is calculated only for $v \in \langle 0, \frac{\pi}{2} \rangle$. Due to the symmetry, the solution in the other quadrants can be easily obtained. With this restriction, the parameter *W*, eq. (2.3), is always non-negative, which simplifies derivations.

The SOS coordinates can be transformed to the Cartesian coordinates with expressions including infinite power series with generalized binomial coefficients [2]. The solution is based on the Lagrange's inversion theorem (see the solution of trinomial equation in [5], [6]; nevertheless, the solution of the problem using infinite series with generalized binomial coefficients was first reported in [7]). The expressions cannot be written in a closed form, nevertheless, they can still be denoted as "analytical".

In order to fulfill the aim of this article, the already derived relations for the SOS coordinates [2] are needed. Therefore, they are listed in the following section. It appeared [2] that the expressions have to be derived sepatately in so called the small-v region and separately in the large-v region, and that the border line between the two regions is defined by the formula

$$W_{border}(R, v) = \sqrt{\frac{\mu^\mu}{(1+\mu)^{1+\mu}}} = constant \quad . \tag{2.4}$$

In Fig. 1, the border line between the two regions is displayed as well. Therefore, the formulae are listed separatelly for the small-v region and for the large-v region in the following sections.

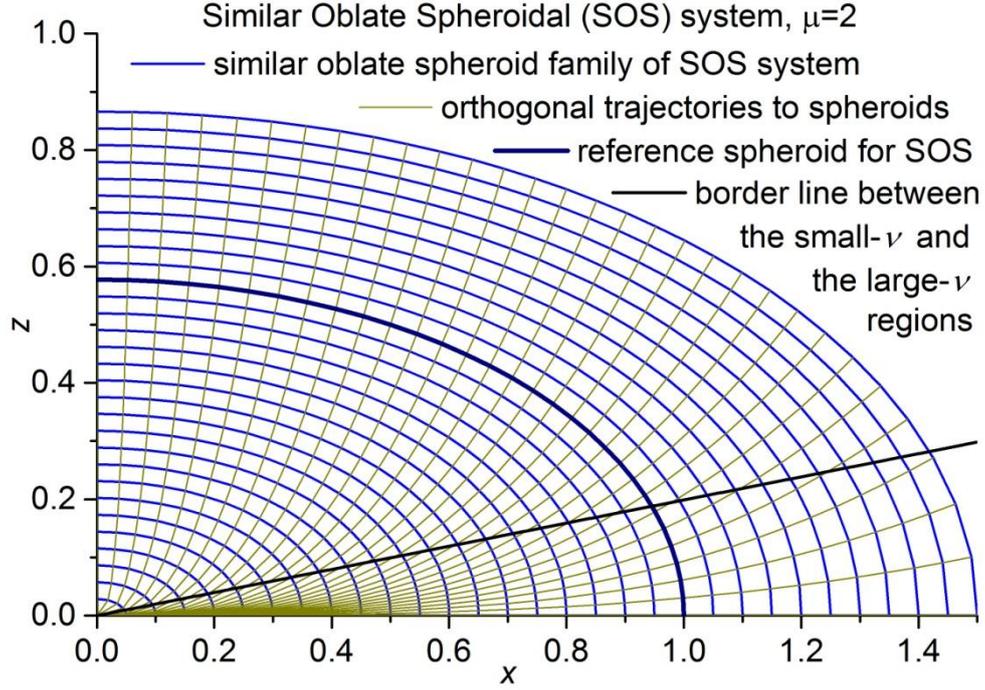

**Figure 1.** One quadrant in *x-z* plane composed of the SOS coordinate system (with *μ*=2) lines around the reference spheroid with the equatorial radius $R_0$=1. Both constant-*R* coordinate surfaces (spheroids, here represented by their sections by *x-z* plane) and orthogonal trajectories to them (power curves) having a constant value of *v* are displayed. The angular spacing of the parametric angle *v* lines is 3°. The boundary line between the small-*v* and the large-*v* regions is displayed as well.

### *2.2 Formulae for the small-v region*

The SOS coordinates *R*, *v*, *λ* transformation to the 3D Cartesian coordinates is as follows [2] in the small-v region:

$$x_{3D} = x(v,R)\cos\lambda = \cos\lambda\ R \sum_{k=0}^{\infty} \binom{-\frac{1}{2}-\mu k}{k} \frac{-\frac{1}{2}}{-\frac{1}{2}-\mu k}(W^2)^k \qquad (2.5)$$

$$y_{3D} = x(v,R)\sin\lambda = \sin\lambda\ R \sum_{k=0}^{\infty} \binom{-\frac{1}{2}-\mu k}{k} \frac{-\frac{1}{2}}{-\frac{1}{2}-\mu k}(W^2)^k \qquad (2.6)$$

$$z_{3D} = z(v,R) = W\frac{R}{\sqrt{1+\mu}}\sum_{k=0}^{\infty} \binom{-\frac{1+\mu}{2}-\mu k}{k} \frac{-\frac{1+\mu}{2}}{-\frac{1+\mu}{2}-\mu k}\left(W^2\right)^k \qquad (2.7)$$

where the parameter *W* was defined in (2.3). The convergence limit for the series in (2.5) – (2.7) is $W<W_{border}$, where $W_{border}$ is given by (2.4). The power series contain generalized binomial coefficients.

Although the transformation using infinite power series may seem to be complicated, it has an advantage of relatively simple differentiability and integration, resulting again in infinite power series. The partial derivatives are [2]:

$$\frac{\partial x_{3D}(v,R,\lambda)}{\partial R} = \cos\lambda \frac{\partial x(v,R)}{\partial R} = \cos\lambda \sum_{k=0}^{\infty} \binom{-\frac{1}{2}-\mu k}{k} (W^2)^k \quad , \tag{2.8}$$

$$\frac{\partial y_{3D}(v,R,\lambda)}{\partial R} = \sin\lambda \frac{\partial x(v,R)}{\partial R} = \sin\lambda \sum_{k=0}^{\infty} \binom{-\frac{1}{2}-\mu k}{k} (W^2)^k \quad , \tag{2.9}$$

$$\frac{\partial z_{3D}(v,R,\lambda)}{\partial R} = \frac{\partial z(v,R)}{\partial R} = W\sqrt{(1+\mu)} \sum_{k=0}^{\infty} \binom{-\frac{\mu+1}{2}-\mu k}{k} (W^2)^k \quad , \tag{2.10}$$

$$\frac{\partial x_{3D}(v,R,\lambda)}{\partial v} = \cos\lambda \frac{\partial x(v,R)}{\partial v} = \cos\lambda \; R \frac{1}{\mu W} \frac{\partial W}{\partial v} \left\{ \sum_{k=0}^{\infty} \binom{-\frac{1}{2}-\mu k}{k} (W^2)^k - \sum_{k=0}^{\infty} \binom{-\frac{1}{2}-\mu k}{k} \frac{-\frac{1}{2}}{-\frac{1}{2}-\mu k} (W^2)^k \right\} \quad , \tag{2.11}$$

$$\frac{\partial y_{3D}(v,R,\lambda)}{\partial v} = \sin\lambda \frac{\partial x(v,R)}{\partial v} = \sin\lambda \; R \frac{1}{\mu W} \frac{\partial W}{\partial v} \left\{ \sum_{k=0}^{\infty} \binom{-\frac{1}{2}-\mu k}{k} (W^2)^k - \sum_{k=0}^{\infty} \binom{-\frac{1}{2}-\mu k}{k} \frac{-\frac{1}{2}}{-\frac{1}{2}-\mu k} (W^2)^k \right\} \quad , \tag{2.12}$$

$$\frac{\partial z_{3D}(v,R)}{\partial v} = \frac{\partial z(v,R)}{\partial v} = \frac{1}{\sqrt{1+\mu}} R \frac{1}{\mu} \frac{\partial W}{\partial v} \left\{ (1+\mu)\sum_{k=0}^{\infty} \binom{-\frac{1+\mu}{2}-\mu k}{k} (W^2)^k - \sum_{k=0}^{\infty} \binom{-\frac{1+\mu}{2}-\mu k}{k} \frac{-\frac{1+\mu}{2}}{-\frac{1+\mu}{2}-\mu k} (W^2)^k \right\} \quad , \tag{2.13}$$

$$\frac{\partial x_{3D}(v,R,\lambda)}{\partial \lambda} = \frac{\partial \cos\lambda}{\partial \lambda} x(v,R) = -\sin\lambda \; R \sum_{k=0}^{\infty} \binom{-\frac{1}{2}-\mu k}{k} \frac{-\frac{1}{2}}{-\frac{1}{2}-\mu k} (W^2)^k \quad , \tag{2.14}$$

$$\frac{\partial y_{3D}(v,R,\lambda)}{\partial \lambda} = \frac{\partial \sin\lambda}{\partial \lambda} x(v,R) = \cos\lambda \; R \sum_{k=0}^{\infty} \binom{-\frac{1}{2}-\mu k}{k} \frac{-\frac{1}{2}}{-\frac{1}{2}-\mu k} (W^2)^k \quad , \tag{2.15}$$

$$\frac{\partial z_{3D}(v,R)}{\partial \lambda} = 0 \quad . \tag{2.16}$$

The metric scale factors can be sumarized as follows [2]:

$$h_R = \sqrt{\sum_{k=0}^{\infty} \binom{-\mu k}{k} (W^2)^k} \quad , \tag{2.17}$$

$$h_v = \frac{R}{\sqrt{1+\mu}} \frac{\partial W}{\partial v} \sqrt{\sum_{k=0}^{\infty} \binom{-(\mu+2)-\mu k}{k} (W^2)^k} \quad , \tag{2.18}$$

$$h_\lambda = x(v,R) = R \sum_{k=0}^{\infty} \binom{-\frac{1}{2}-\mu k}{k} \frac{-\frac{1}{2}}{-\frac{1}{2}-\mu k} (W^2)^k \quad . \tag{2.19}$$

Here

$$\frac{\partial W}{\partial v} = \left(\frac{R}{R_0}\right)^\mu \frac{1+\mu \sin^2 v}{\cos^{2+\mu} v} \quad . \tag{2.20}$$

To complete the list of the already derived relations [2] for the SOS coordinate system, also the Jacobian determinant in the small-v region is reported,

$$\Im = h_R h_\nu h_\lambda = \frac{R^2}{\sqrt{1+\mu}} \frac{\partial W}{\partial \nu} \sum_{k=0}^{\infty} \binom{-\frac{\mu+3}{2}-\mu k}{k} \left(W^2\right)^k \quad , \tag{2.21}$$

as well as the Jacobian divided by the square of the $h_R$ scale factor

$$\frac{\Im}{h_R^2} = \frac{R^2}{\sqrt{1+\mu}} \frac{\partial W}{\partial \nu} \sum_{k=0}^{\infty} \frac{-\frac{\mu+3}{2}}{-\frac{\mu+3}{2}-\mu k} \binom{-\frac{\mu+3}{2}-\mu k}{k} \left(W^2\right)^k \quad . \tag{2.22}$$

## 2.3 Formulae for the large-v region

3D SOS coordinates transformation to the Cartesian coordinates in the large-$v$ region is as follows [2]:

$$x_{3D} = x(\nu,R)\cos\lambda = \cos\lambda \, W^{-\frac{1}{1+\mu}} R \sum_{k=0}^{\infty} \binom{-\frac{1}{2(1+\mu)}+\frac{\mu}{1+\mu}k}{k} \frac{-\frac{1}{2(1+\mu)}}{-\frac{1}{2(1+\mu)}+\frac{\mu}{1+\mu}k} \left(W^{-\frac{2}{1+\mu}}\right)^k \quad , \tag{2.23}$$

$$y_{3D} = x(\nu,R)\sin\lambda = \sin\lambda \, W^{-\frac{1}{1+\mu}} R \sum_{k=0}^{\infty} \binom{-\frac{1}{2(1+\mu)}+\frac{\mu}{1+\mu}k}{k} \frac{-\frac{1}{2(1+\mu)}}{-\frac{1}{2(1+\mu)}+\frac{\mu}{1+\mu}k} \left(W^{-\frac{2}{1+\mu}}\right)^k \quad , \tag{2.24}$$

$$z_{3D} = z(\nu,R) = \frac{1}{\sqrt{1+\mu}} R \sum_{k=0}^{\infty} \binom{-\frac{1}{2}+\frac{\mu}{1+\mu}k}{k} \frac{-\frac{1}{2}}{-\frac{1}{2}+\frac{\mu}{1+\mu}k} \left(W^{-\frac{2}{1+\mu}}\right)^k \quad , \tag{2.25}$$

where the parameter $W$ equals to (2.3). The convergence limit for the series in (2.23) – (2.25) is $W > W_{border}$, where $W_{border}$ is given by (2.4).

From (2.5) and (2.7), as well as from (2.23) and (2.25), it follows that the ratio of $x$ and $z$ coordinates in the $x$-$z$ plane is dependent only on the parameter $W$. It means that, for $W$=constant while changing $R$, the points ($x$, $z$) lie on a straight line starting at zero, regardless the particular value of $R$ and $v$. In 3D, it is a cone instead. This is valid for both the small-$v$ region and the large-$v$ region, and even in any vicinity of the convergence limit $W=W_{border}$. As a consequence of this, also the border between the small- and the large-$v$ regions (characterized by $W_{border}$ being constant according to (2.4)), is a straight line starting at zero, regardless if the series in (2.5) – (2.7) and (2.23) – (2.25) converge or not at the border line.

The partial derivatives in the large-$v$ region are:

$$\frac{\partial x_{3D}(\nu,R,\lambda)}{\partial R} = \cos\lambda \frac{\partial x(\nu,R)}{\partial R} = \cos\lambda \frac{W^{-\frac{1}{1+\mu}}}{1+\mu} \sum_{k=0}^{\infty} \binom{-\frac{1}{2}\frac{1}{1+\mu}+\frac{\mu}{1+\mu}k}{k} \left(W^{-\frac{2}{1+\mu}}\right)^k \quad , \tag{2.26}$$

$$\frac{\partial y_{3D}(\nu,R,\lambda)}{\partial R} = \sin\lambda \frac{\partial x(\nu,R)}{\partial R} = \sin\lambda \frac{W^{-\frac{1}{1+\mu}}}{1+\mu} \sum_{k=0}^{\infty} \binom{-\frac{1}{2}\frac{1}{1+\mu}+\frac{\mu}{1+\mu}k}{k} \left(W^{-\frac{2}{1+\mu}}\right)^k \quad , \tag{2.27}$$

$$\frac{\partial z_{3D}(\nu,R,\lambda)}{\partial R} = \frac{\partial z(\nu,R)}{\partial R} = \frac{1}{\sqrt{1+\mu}} \sum_{k=0}^{\infty} \binom{-\frac{1}{2}+\frac{\mu}{1+\mu}k}{k} \left(W^{-\frac{2}{1+\mu}}\right)^k \quad , \tag{2.28}$$

$$\frac{\partial x_{3D}(v,R,\lambda)}{\partial v} = \cos\lambda \frac{\partial x(v,R)}{\partial v} = \cos\lambda\, R\, \frac{1}{\mu W^{\frac{2+\mu}{1+\mu}}} \frac{\partial W}{\partial v} \left\{ \frac{1}{1+\mu} \sum_{k=0}^{\infty} \binom{-\frac{1}{2}\frac{1}{1+\mu}+\frac{\mu}{1+\mu}k}{k} \left(W^{-\frac{2}{1+\mu}}\right)^k - \sum_{k=0}^{\infty} \binom{-\frac{1}{2}\frac{1}{1+\mu}+\frac{\mu}{1+\mu}k}{k} \frac{-\frac{1}{2}\frac{1}{1+\mu}}{-\frac{1}{2}\frac{1}{1+\mu}+\frac{\mu}{1+\mu}k} \left(W^{-\frac{2}{1+\mu}}\right)^k \right\},$$
(2.29)

$$\frac{\partial y_{3D}(v,R,\lambda)}{\partial v} = \sin\lambda \frac{\partial x(v,R)}{\partial v} = \sin\lambda\, R\, \frac{1}{\mu W^{\frac{2+\mu}{1+\mu}}} \frac{\partial W}{\partial v} \left\{ \frac{1}{1+\mu} \sum_{k=0}^{\infty} \binom{-\frac{1}{2}\frac{1}{1+\mu}+\frac{\mu}{1+\mu}k}{k} \left(W^{-\frac{2}{1+\mu}}\right)^k - \sum_{k=0}^{\infty} \binom{-\frac{1}{2}\frac{1}{1+\mu}+\frac{\mu}{1+\mu}k}{k} \frac{-\frac{1}{2}\frac{1}{1+\mu}}{-\frac{1}{2}\frac{1}{1+\mu}+\frac{\mu}{1+\mu}k} \left(W^{-\frac{2}{1+\mu}}\right)^k \right\},$$
(2.30)

$$\frac{\partial z_{3D}(v,R)}{\partial v} = \frac{\partial z(v,R)}{\partial v} = \frac{1}{\sqrt{1+\mu}} R\, \frac{1}{\mu W} \frac{\partial W}{\partial v} \left\{ \sum_{k=0}^{\infty} \binom{-\frac{1}{2}+\frac{\mu}{1+\mu}k}{k} \left(W^{-\frac{2}{1+\mu}}\right)^k - \sum_{k=0}^{\infty} \binom{-\frac{1}{2}+\frac{\mu}{1+\mu}k}{k} \frac{-\frac{1}{2}}{-\frac{1}{2}+\frac{\mu}{1+\mu}k} \left(W^{-\frac{2}{1+\mu}}\right)^k \right\}, \quad (2.31)$$

$$\frac{\partial x_{3D}(v,R,\lambda)}{\partial \lambda} = \frac{\partial \cos\lambda}{\partial \lambda} x(v,R) = -\sin\lambda\, W^{-\frac{1}{1+\mu}} R \sum_{k=0}^{\infty} \binom{-\frac{1}{2}\frac{1}{1+\mu}+\frac{\mu}{1+\mu}k}{k} \frac{-\frac{1}{2}\frac{1}{1+\mu}}{-\frac{1}{2}\frac{1}{1+\mu}+\frac{\mu}{1+\mu}k} \left(W^{-\frac{2}{1+\mu}}\right)^k, \quad (2.32)$$

$$\frac{\partial y_{3D}(v,R,\lambda)}{\partial \lambda} = \frac{\partial \sin\lambda}{\partial \lambda} x(v,R) = \cos\lambda\, W^{-\frac{1}{1+\mu}} R \sum_{k=0}^{\infty} \binom{-\frac{1}{2}\frac{1}{1+\mu}+\frac{\mu}{1+\mu}k}{k} \frac{-\frac{1}{2}\frac{1}{1+\mu}}{-\frac{1}{2}\frac{1}{1+\mu}+\frac{\mu}{1+\mu}k} \left(W^{-\frac{2}{1+\mu}}\right)^k, \quad (2.33)$$

$$\frac{\partial z_{3D}(v,R)}{\partial \lambda} = 0\;. \quad (2.34)$$

The metric scale factors can be sumarized as follows:

$$h_R = \frac{1}{\sqrt{1+\mu}} \sqrt{\sum_{k=0}^{\infty} \binom{\frac{\mu}{1+\mu}k}{k} \left(W^{-\frac{2}{1+\mu}}\right)^k}\;, \quad (2.35)$$

$$h_v = \frac{R}{1+\mu} W^{-\frac{2+\mu}{1+\mu}} \frac{\partial W}{\partial v} \sqrt{\sum_{k=0}^{\infty} \binom{-\frac{2+\mu}{1+\mu}+\frac{\mu}{1+\mu}k}{k} \left(W^{-\frac{2}{1+\mu}}\right)^k}\;, \quad (2.36)$$

$$h_\lambda = x(v,R) = W^{-\frac{1}{1+\mu}} R \sum_{k=0}^{\infty} \binom{-\frac{1}{2}\frac{1}{1+\mu}+\frac{\mu}{1+\mu}k}{k} \frac{-\frac{1}{2}\frac{1}{1+\mu}}{-\frac{1}{2}\frac{1}{1+\mu}+\frac{\mu}{1+\mu}k} \left(W^{-\frac{2}{1+\mu}}\right)^k\;. \quad (2.37)$$

Finally, also the Jacobian determinant for the large-$v$ region is reported,

$$\mathfrak{I} = h_R h_v h_\lambda = \frac{R^2 W^{-\frac{3+\mu}{1+\mu}}}{\sqrt{(1+\mu)^3}} \frac{\partial W}{\partial v} \sum_{k=0}^{\infty} \binom{-\frac{1}{2}\frac{\mu+3}{1+\mu}+\frac{\mu}{1+\mu}k}{k} \left(W^{-\frac{2}{1+\mu}}\right)^k, \quad (2.38)$$

as well as the Jacobian divided by the square of the $h_R$ scale factor

$$\frac{\mathfrak{I}}{h_R^2} = \frac{R^2}{\sqrt{1+\mu}} W^{-\frac{3+\mu}{1+\mu}} \frac{\partial W}{\partial v} \sum_{k=0}^{\infty} \frac{-\frac{1}{2}\frac{\mu+3}{\mu+1}}{-\frac{1}{2}\frac{\mu+3}{1+\mu}+\frac{\mu}{1+\mu}k} \binom{-\frac{1}{2}\frac{\mu+3}{1+\mu}+\frac{\mu}{1+\mu}k}{k} \left(W^{-\frac{2}{1+\mu}}\right)^k\;. \quad (2.39)$$

## 3 Usefull combinatorial identities

Operations with the above reported series is not as straightforward as a simple multiplication or summation of two quantities. Nevertheless, it is possible to deal with them with a help of known combinatorial identities [7], [8], [9], [10]. The important ones, used in the derivations, are listed below.

Pólya and Szegö identities (see [7], or [8], No. 1.120 and 1.121, or [11]) are

$$\frac{p^{a+1}}{(1-b)p+b} = \sum_{k=0}^{\infty} \binom{a+bk}{k} r^k \quad , \tag{3.1}$$

$$p^a = \sum_{k=0}^{\infty} \frac{a}{a+bk} \binom{a+bk}{k} r^k \quad , \tag{3.2}$$

where the following equation is fulfilled

$$-rp^b + p = 1 \quad , \tag{3.3}$$

and the series have the convergence limit

$$|r| < \left|\frac{(b-1)^{b-1}}{b^b}\right| \tag{3.4}$$

(for the discussion on the convergence limit, see [2], [12]).

Hagen-Rothe [9] identity for generalized binomial coefficients is

$$\sum_{m=0}^{k} \frac{a}{a+bm} \binom{a+bm}{m}\binom{c+b(k-m)}{k-m} = \binom{a+c+bk}{k} \quad , \tag{3.5}$$

and Jensen (see [8], No. 3.143) identity reads

$$\sum_{m=0}^{k} \binom{a+bm}{m}\binom{c+b(k-m)}{k-m} = \sum_{m=0}^{k} \binom{a+t+bm}{m}\binom{c-t+b(k-m)}{k-m} \quad . \tag{3.6}$$

Cauchy product for the series with generalized binomial coefficients is

$$\left[\sum_{k=0}^{\infty} \frac{a}{a+bk}\binom{a+bk}{k}r^k\right]\left[\sum_{k=0}^{\infty}\binom{c+bk}{k}r^k\right] = \sum_{k=0}^{\infty} r^k \sum_{m=0}^{k} \frac{a}{a+bm}\binom{a+bm}{m}\binom{c+b(k-m)}{k-m}, \tag{3.7}$$

or similarly

$$\left[\sum_{k=0}^{\infty} \binom{a+bk}{k}r^k\right]\left[\sum_{k=0}^{\infty}\binom{c+bk}{k}r^k\right] = \sum_{k=0}^{\infty} r^k \sum_{m=0}^{k} \binom{a+bm}{m}\binom{c+b(k-m)}{k-m} \quad , \tag{3.8}$$

By combining the Hagen-Rothe identity (3.5) with the first Cauchy product (3.7), the following identity is obtained:

$$\left[\sum_{k=0}^{\infty} \frac{a}{a+bk}\binom{a+bk}{k}r^k\right]\left[\sum_{k=0}^{\infty}\binom{c+bk}{k}r^k\right] = \sum_{k=0}^{\infty}\binom{a+c+bk}{k}r^k \quad . \tag{3.9}$$

From it also follows

$$\sum_{k=0}^{\infty} \frac{a}{a+bk} \binom{a+bk}{k} r^k = \frac{\sum_{k=0}^{\infty} \binom{a+bk}{k} r^k}{\sum_{k=0}^{\infty} \binom{bk}{k} r^k} \quad . \tag{3.10}$$

By combining the Jensen identity (3.6) with the second type Cauchy product (3.8), the following identity is obtained:

$$\left[\sum_{k=0}^{\infty} \binom{a+bk}{k} r^k\right]\left[\sum_{k=0}^{\infty} \binom{c+bk}{k} r^k\right] = \left[\sum_{k=0}^{\infty} \binom{a+t+bk}{k} r^k\right]\left[\sum_{k=0}^{\infty} \binom{c-t+bk}{k} r^k\right] \quad , \tag{3.11}$$

For generalized binomial coefficients, also the following identities hold [8]:

$$\binom{\beta+k}{k} \frac{1}{\beta+k} = \binom{\beta+k-1}{k} \frac{1}{\beta} = (-1)^{k+1} \binom{-\beta}{k} \frac{1}{-\beta} \quad . \tag{3.12}$$

Thus, there are three equivalent forms for the quantity expressed by generalized binomial coefficients. In order to complete the set of binomial identities, the following are also listed [8]:

$$\binom{\alpha}{k} = \binom{\alpha}{k-1} \frac{\alpha-k+1}{k} \quad , \tag{3.13}$$

$$\binom{\alpha+1}{k} - \binom{\alpha}{k} = \binom{\alpha}{k-1} \quad , \tag{3.14}$$

where $\alpha$ is a real number. The binomial theorem will be used as well:

$$(1+r)^\alpha = \sum_{k=0}^{\infty} \binom{\alpha}{k} r^k \quad . \tag{3.15}$$

Finally, the generating function of the central binomial coefficients, which has a closed form

$$\sum_{k=0}^{\infty} \binom{2k}{k} r^k = \frac{1}{\sqrt{1-4r}} \quad \text{where} \quad |r| < \frac{1}{4} \tag{3.16}$$

is to be mentioned [10], together with its more general counterpart

$$\sum_{k=0}^{\infty} \binom{a+2k}{k} r^k = \frac{1}{\sqrt{1-4r}} \left(\frac{1-\sqrt{1-4r}}{2r}\right)^a \quad \text{where} \quad |r| < \frac{1}{4} \quad . \tag{3.17}$$

## 4 Unit vectors in SOS coordinate system for the small-*v* region

In this chapter, the unit vectors are derived in the small-v region.

### 4.1 Unit vector in R direction

The norm in the case of *R*-direction is (see (2.8) – (2.10))

$$\sqrt{\left(\frac{\partial x_{3D}(v,R,\lambda)}{\partial R}\right)^2 + \left(\frac{\partial y_{3D}(v,R,\lambda)}{\partial R}\right)^2 + \left(\frac{\partial z_{3D}(v,R)}{\partial R}\right)^2} = \sqrt{\cos^2\lambda \left(\frac{\partial x(v,R)}{\partial R}\right)^2 + \sin^2\lambda \left(\frac{\partial x(v,R)}{\partial R}\right)^2 + \left(\frac{\partial z(v,R)}{\partial R}\right)^2} =$$

$$\sqrt{\left(\frac{\partial x(v,R)}{\partial R}\right)^2 + \left(\frac{\partial z(v,R)}{\partial R}\right)^2} = h_R \quad , \tag{4.1}$$

and thus the unit vector is

$$\widehat{R} = \begin{pmatrix} \frac{\partial x_{3D}(v,R,\lambda)}{\partial R}/h_R \\ \frac{\partial y_{3D}(v,R,\lambda)}{\partial R}/h_R \\ \frac{\partial z_{3D}(v,R)}{\partial R}/h_R \end{pmatrix} = \begin{pmatrix} \cos\lambda \frac{\partial x(v,R)}{\partial R}/h_R \\ \sin\lambda \frac{\partial x(v,R)}{\partial R}/h_R \\ \frac{\partial z(v,R)}{\partial R}/h_R \end{pmatrix} \qquad (4.2)$$

First, the component containing derivative of *x* with respect to *R* is derived, see (2.8), (2.9) and (2.17):

$$\frac{\partial x(v,R)}{\partial R}/h_R = \frac{\sum_{k=0}^{\infty}\binom{-\frac{1}{2}-\mu k}{k}(W^2)^k}{\sqrt{\sum_{k=0}^{\infty}\binom{-\mu k}{k}(W^2)^k}} \qquad (4.3)$$

When Cauchy product combined with Jensen identity (3.11) are used for the numerator (with $b=-\mu$, $a=c=-½$, $t=½$), then

$$\sum_{k=0}^{\infty}\binom{-\frac{1}{2}-\mu k}{k}(W^2)^k = \sqrt{\sum_{k=0}^{\infty}\binom{-\frac{1}{2}-\mu k}{k}(W^2)^k \sum_{k=0}^{\infty}\binom{-\frac{1}{2}-\mu k}{k}(W^2)^k} = \sqrt{\sum_{k=0}^{\infty}\binom{-\mu k}{k}(W^2)^k \sum_{k=0}^{\infty}\binom{-1-\mu k}{k}(W^2)^k} \qquad (4.4)$$

and thus

$$\frac{\partial x(v,R)}{\partial R}/h_R = \frac{\sqrt{\sum_{k=0}^{\infty}\binom{-\mu k}{k}(W^2)^k}\sqrt{\sum_{k=0}^{\infty}\binom{-1-\mu k}{k}(W^2)^k}}{\sqrt{\sum_{k=0}^{\infty}\binom{-\mu k}{k}(W^2)^k}} = \sqrt{\sum_{k=0}^{\infty}\binom{-1-\mu k}{k}(W^2)^k} \qquad (4.5)$$

Similarly for *z* coordinate:

$$\frac{\partial z(v,R)}{\partial R}/h_R = \frac{W\sqrt{1+\mu}\sum_{k=0}^{\infty}\binom{-\frac{\mu+1}{2}-\mu k}{k}(W^2)^k}{\sqrt{\sum_{k=0}^{\infty}\binom{-\mu k}{k}(W^2)^k}} \qquad (4.6)$$

and using (3.11) for the series in numerator (with $b=-\mu$, $a=c=-(\mu+1)/2$, $t=(\mu+1)/2$), the following is obtained:

$$\frac{\partial z(v,R)}{\partial R}/h_R = \frac{W\sqrt{1+\mu}\sqrt{\sum_{k=0}^{\infty}\binom{-\mu k}{k}(W^2)^k}\sqrt{\sum_{k=0}^{\infty}\binom{-(1+\mu)-\mu k}{k}(W^2)^k}}{\sqrt{\sum_{k=0}^{\infty}\binom{-\mu k}{k}(W^2)^k}} = W\sqrt{1+\mu}\sqrt{\sum_{k=0}^{\infty}\binom{-(1+\mu)-\mu k}{k}(W^2)^k} \qquad (4.7)$$

The unit vector $\widehat{R}$ (4.2) is then

$$\widehat{R} = \begin{pmatrix} \cos\lambda \sqrt{\sum_{k=0}^{\infty}\binom{-1-\mu k}{k}(W^2)^k} \\ \sin\lambda \sqrt{\sum_{k=0}^{\infty}\binom{-1-\mu k}{k}(W^2)^k} \\ W\sqrt{1+\mu} \sqrt{\sum_{k=0}^{\infty}\binom{-(1+\mu)-\mu k}{k}(W^2)^k} \end{pmatrix} \qquad (4.8)$$

For the special case when $\mu=0$ (i.e. the case equivalent to the spherical coordinates), the parameter $W$, see (2.3), is simplified to

$$W(\mu = 0) = \frac{\sin \nu}{\cos \nu} \quad , \tag{4.9}$$

and the result is (thanks to the binomial theorem (3.15))

$$\sqrt{\sum_{k=0}^{\infty}\binom{-1-\mu k}{k}(W^2)^k} = \sqrt{\sum_{k=0}^{\infty}\binom{-1}{k}(W^2)^k} = \sqrt{(1+W^2)^{-1}} = \sqrt{\frac{1}{1+\frac{\sin^2\nu}{\cos^2\nu}}} = \cos \nu \quad . \tag{4.10}$$

The unit vector for the special case of spherical-like coordinates ($\mu=0$) is thus

$$\widehat{R}(\mu = 0) = \begin{pmatrix} \cos \lambda \cos \nu \\ \sin \lambda \cos \nu \\ \sin \nu \end{pmatrix} \quad , \tag{4.11}$$

which is correct for the equivalent of the spherical coordinates assuming that $\nu$ is not a polar angle but increases when moving from the equator towards the pole.

Another partial confirmation, that the above derivation of the unit vector is correct, is that the unit vector length

$$\left[\left(\cos \lambda \sqrt{\sum_{k=0}^{\infty}\binom{-1-\mu k}{k}(W^2)^k}\right)^2 + \left(\sin \lambda \sqrt{\sum_{k=0}^{\infty}\binom{-1-\mu k}{k}(W^2)^k}\right)^2 + \left(W\sqrt{1+\mu}\sqrt{\sum_{k=0}^{\infty}\binom{-(1+\mu)-\mu k}{k}(W^2)^k}\right)^2\right]^{1/2} \tag{4.12}$$

has to be equal to one. The length square can be simplified to

$$\sum_{k=0}^{\infty}\binom{-1-\mu k}{k}(W^2)^k + W^2(1+\mu) \sum_{k=0}^{\infty}\binom{-(1+\mu)-\mu k}{k}(W^2)^k \quad . \tag{4.13}$$

When (3.1) is used with $r = W^2$, $b = -\mu$, and $a = -1$ in the first term while $a = -(\mu+1)$ in the second term, the relation can be rewritten to the form

$$\frac{p^{-1+1}}{(1+\mu)p-\mu} + W^2(1+\mu)\frac{p^{-(1+\mu)+1}}{(1+\mu)p-\mu} = \frac{1}{(1+\mu)p-\mu} + W^2(1+\mu)\frac{p^{-\mu}}{(1+\mu)p-\mu} \quad . \tag{4.14}$$

According to (3.3)

$$W^2 p^{-\mu} = p - 1 \quad , \tag{4.15}$$

and thus further simplification follows:

$$\frac{1}{(1+\mu)p-\mu} + (1+\mu)\frac{p-1}{(1+\mu)p-\mu} = \frac{1+p-1+\mu p-\mu}{p+\mu p-\mu} = 1 \quad . \tag{4.16}$$

Indeed, the length of the unit vector is equal to one.

### 4.2  Unit vector in v direction

The norm in this case is

$$\sqrt{\left(\frac{\partial x_{3D}(v,R,\lambda)}{\partial v}\right)^2 + \left(\frac{\partial y_{3D}(v,R,\lambda)}{\partial v}\right)^2 + \left(\frac{\partial z_{3D}(v,R)}{\partial v}\right)^2} = \sqrt{\cos^2 \lambda \left(\frac{\partial x(v,R)}{\partial v}\right)^2 + \sin^2 \lambda \left(\frac{\partial x(v,R)}{\partial v}\right)^2 + \left(\frac{\partial z(v,R)}{\partial v}\right)^2}$$
$$= \sqrt{\left(\frac{\partial x(v,R)}{\partial v}\right)^2 + \left(\frac{\partial z(v,R)}{\partial v}\right)^2} = h_v \quad . \quad (4.17)$$

And, therefore, the unit vector is equal to

$$\hat{\boldsymbol{v}} = \begin{pmatrix} \frac{\partial x_{3D}(v,R,\lambda)}{\partial v}/h_v \\ \frac{\partial y_{3D}(v,R,\lambda)}{\partial v}/h_v \\ \frac{\partial z_{3D}(v,R)}{\partial v}/h_v \end{pmatrix} = \begin{pmatrix} \cos \lambda \frac{\partial x(v,R)}{\partial v}/h_v \\ \sin \lambda \frac{\partial x(v,R)}{\partial v}/h_v \\ \frac{\partial z(v,R)}{\partial v}/h_v \end{pmatrix} \quad . \quad (4.18)$$

The first component of the unit vector $\hat{\boldsymbol{v}}$ is thus proportional to (see (2.11), (2.12) and (2.18))

$$\frac{\partial x(v,R)}{\partial v}/h_v = \frac{R\frac{1}{\mu W}\frac{\partial W}{\partial v}\left\{\sum_{k=0}^{\infty}\binom{-\frac{1}{2}-\mu k}{k}(W^2)^k - \sum_{k=0}^{\infty}\binom{-\frac{1}{2}-\mu k}{k}\frac{-\frac{1}{2}}{-\frac{1}{2}-\mu k}(W^2)^k\right\}}{\frac{R}{\sqrt{1+\mu}}\frac{\partial W}{\partial v}\sqrt{\sum_{k=0}^{\infty}\binom{-\mu k-(\mu+2)}{k}(W^2)^k}} =$$

$$\frac{\sqrt{1+\mu}}{\mu W}\left\{\frac{\sum_{k=0}^{\infty}\binom{-\frac{1}{2}-\mu k}{k}(W^2)^k}{\sqrt{\sum_{k=0}^{\infty}\binom{-\mu k-(\mu+2)}{k}(W^2)^k}} - \frac{\sum_{k=0}^{\infty}\binom{-\frac{1}{2}-\mu k}{k}\frac{-\frac{1}{2}}{-\frac{1}{2}-\mu k}(W^2)^k}{\sqrt{\sum_{k=0}^{\infty}\binom{-\mu k-(\mu+2)}{k}(W^2)^k}}\right\} \quad . \quad (4.19)$$

When Cauchy product combined with Jensen identity (3.11) are used for the first-term numerator (with b= −μ, a=c= −½, t= (μ +3/2)), the following is found (similarly as in the derivation of the unit R vector),

$$\sum_{k=0}^{\infty}\binom{-\frac{1}{2}-\mu k}{k}(W^2)^k = \sqrt{\sum_{k=0}^{\infty}\binom{-\frac{1}{2}-\mu k}{k}(W^2)^k \sum_{k=0}^{\infty}\binom{-\frac{1}{2}-\mu k}{k}(W^2)^k} =$$
$$\sqrt{\sum_{k=0}^{\infty}\binom{-(\mu+2)-\mu k}{k}(W^2)^k \sum_{k=0}^{\infty}\binom{(\mu+1)-\mu k}{k}(W^2)^k} \quad , \quad (4.20)$$

and thus

$$\frac{\partial x(v,R)}{\partial v}/h_v = \frac{\sqrt{1+\mu}}{\mu W}\left\{\sqrt{\sum_{k=0}^{\infty}\binom{(\mu+1)-\mu k}{k}(W^2)^k} - \frac{\sum_{k=0}^{\infty}\binom{-\frac{1}{2}-\mu k}{k}\frac{-\frac{1}{2}}{-\frac{1}{2}-\mu k}(W^2)^k}{\sqrt{\sum_{k=0}^{\infty}\binom{-\mu k-(\mu+2)}{k}(W^2)^k}}\right\} \quad . \quad (4.21)$$

As the identity (3.10) is valid, and as also the following identity holds thanks to the Cauchy product combined with Jensen identity (3.11), with b= −μ, a=c= 0, t=(μ+1),

$$\sqrt{\sum_{k=0}^{\infty}\binom{-\mu k}{k}(W^2)^k \sum_{k=0}^{\infty}\binom{-\mu k}{k}(W^2)^k} = \sqrt{\sum_{k=0}^{\infty}\binom{-\mu-1-\mu k}{k}(W^2)^k \sum_{k=0}^{\infty}\binom{\mu+1-\mu k}{k}(W^2)^k} \quad ,$$
$$(4.22)$$

then the second term in (4.21) is

$$\frac{\sum_{k=0}^{\infty}\binom{-\frac{1}{2}-\mu k}{k}\frac{-\frac{1}{2}}{-\frac{1}{2}-\mu k}(W^2)^k}{\sqrt{\sum_{k=0}^{\infty}\binom{-\mu k-(\mu+2)}{k}(W^2)^k}} = \frac{\frac{\sum_{k=0}^{\infty}\binom{-\frac{1}{2}-\mu k}{k}(W^2)^k}{\sum_{k=0}^{\infty}\binom{-\mu k}{k}(W^2)^k}}{\sqrt{\sum_{k=0}^{\infty}\binom{-\mu k-(\mu+2)}{k}(W^2)^k}} =$$

$$\frac{\sum_{k=0}^{\infty}\binom{-\frac{1}{2}-\mu k}{k}(W^2)^k}{\sqrt{\sum_{k=0}^{\infty}\binom{-\mu k}{k}(W^2)^k \sum_{k=0}^{\infty}\binom{-\mu k}{k}(W^2)^k}\sqrt{\sum_{k=0}^{\infty}\binom{-\mu k-(\mu+2)}{k}(W^2)^k}} =$$

$$\frac{\sum_{k=0}^{\infty}\binom{-\frac{1}{2}-\mu k}{k}(W^2)^k}{\sqrt{\sum_{k=0}^{\infty}\binom{-\mu-1-\mu k}{k}(W^2)^k \sum_{k=0}^{\infty}\binom{\mu+1-\mu k}{k}(W^2)^k}\sqrt{\sum_{k=0}^{\infty}\binom{-\mu k-(\mu+2)}{k}(W^2)^k}} =$$

$$\frac{\sum_{k=0}^{\infty}\binom{-\frac{1}{2}-\mu k}{k}(W^2)^k}{\sqrt{\sum_{k=0}^{\infty}\binom{-\mu-1-\mu k}{k}(W^2)^k}\sqrt{\sum_{k=0}^{\infty}\binom{\mu+1-\mu k}{k}(W^2)^k \sum_{k=0}^{\infty}\binom{-\mu k-(\mu+2)}{k}(W^2)^k}} =$$

$$\frac{\sum_{k=0}^{\infty}\binom{-\frac{1}{2}-\mu k}{k}(W^2)^k}{\sqrt{\sum_{k=0}^{\infty}\binom{-\mu-1-\mu k}{k}(W^2)^k}\sqrt{\sum_{k=0}^{\infty}\binom{-\frac{1}{2}-\mu k}{k}(W^2)^k \sum_{k=0}^{\infty}\binom{-\frac{1}{2}-\mu k}{k}(W^2)^k}} = \frac{1}{\sqrt{\sum_{k=0}^{\infty}\binom{-\mu-1-\mu k}{k}(W^2)^k}} \quad . \quad (4.23)$$

In the second part, the relation (3.11) is again used with b= –μ, a=(μ+1), c= –(μ+2), t= –(μ+3/2) in the denominator. Therefore

$$\left.\frac{\partial x(v,R)}{\partial v}\right/h_v = \frac{\sqrt{1+\mu}}{\mu W}\left\{\sqrt{\sum_{k=0}^{\infty}\binom{(\mu+1)-\mu k}{k}(W^2)^k} - \frac{1}{\sqrt{\sum_{k=0}^{\infty}\binom{-\mu-1-\mu k}{k}(W^2)^k}}\right\} \quad . \quad (4.24)$$

As Pólya and Szegö identity (3.1) holds with $r = W^2$, b= –μ, and as a= μ+1 in the first term while it is a= –μ–1 in the denominator of the second term, then

$$\left.\frac{\partial x(v,R)}{\partial v}\right/h_v = \frac{\sqrt{1+\mu}}{\mu W}\left\{\sqrt{\frac{p^{(\mu+1)+1}}{(1+\mu)p-\mu}} - \frac{1}{\sqrt{\frac{p^{-(\mu+1)+1}}{(1+\mu)p-\mu}}}\right\} = \frac{\sqrt{1+\mu}}{\mu W}\left\{\sqrt{\frac{p^{\mu+2}}{(1+\mu)p-\mu}} - \frac{1}{\sqrt{\frac{p^{-\mu}}{(1+\mu)p-\mu}}}\right\} =$$

$$\frac{\sqrt{1+\mu}}{\mu W}\left\{\frac{p\sqrt{p^\mu}}{\sqrt{(1+\mu)p-\mu}} - \sqrt{p^\mu}\sqrt{(1+\mu)p-\mu}\right\} = \frac{\sqrt{1+\mu}}{\mu W}\frac{p\sqrt{p^\mu}-\sqrt{p^\mu}[(1+\mu)p-\mu]}{\sqrt{(1+\mu)p-\mu}} = \frac{\sqrt{1+\mu}}{\mu W}\sqrt{p^\mu}\frac{p-(1+\mu)p+\mu}{\sqrt{(1+\mu)p-\mu}} =$$

$$\frac{\sqrt{1+\mu}}{\mu W}\sqrt{p^\mu}\frac{-\mu(p-1)}{\sqrt{(1+\mu)p-\mu}} = -\frac{\sqrt{1+\mu}}{W}\sqrt{p^\mu}\frac{p-1}{\sqrt{(1+\mu)p-\mu}} \quad . \quad (4.25)$$

As (see (3.3), assuming $r = W^2$, b= –μ)

$$W^2 = \frac{p-1}{p^{-\mu}} = (p-1)p^\mu \quad \Rightarrow \quad p^\mu = \frac{W^2}{p-1} \quad , \quad (4.26)$$

and thus (using again (3.1), once with a=0 while second time with a= –1)

$$\left.\frac{\partial x(v,R)}{\partial v}\right/h_v = -\frac{\sqrt{1+\mu}}{W}\sqrt{\frac{W^2}{p-1}}\frac{p-1}{\sqrt{(1+\mu)p-\mu}} = -\sqrt{1+\mu}\frac{\sqrt{p-1}}{\sqrt{(1+\mu)p-\mu}} = -\sqrt{1+\mu}\sqrt{\frac{p}{(1+\mu)p-\mu} - \frac{1}{(1+\mu)p-\mu}} =$$

$$-\sqrt{1+\mu}\sqrt{\sum_{k=0}^{\infty}\binom{-\mu k}{k}(W^2)^k - \sum_{k=0}^{\infty}\binom{-1-\mu k}{k}(W^2)^k} = -\sqrt{1+\mu}\sqrt{\sum_{k=0}^{\infty}\left[\binom{-\mu k}{k}-\binom{-1-\mu k}{k}\right](W^2)^k} =$$

$$-\sqrt{1+\mu}\sqrt{\sum_{k=1}^{\infty}\left[\binom{-\mu k}{k}-\binom{-1-\mu k}{k}\right](W^2)^k} \quad . \quad (4.27)$$

Due to the identity (3.14) with α = –1–μk, it follows that

$$\frac{\partial x(v,R)}{\partial v}/h_v = -\sqrt{1+\mu}\sqrt{\sum_{k=1}^{\infty}\binom{-1-\mu k}{k-1}(W^2)^k} = -\sqrt{1+\mu}\sqrt{\sum_{M=0}^{\infty}\binom{-1-\mu(M+1)}{M}(W^2)^{M+1}} =$$
$$-W\sqrt{1+\mu}\sqrt{\sum_{M=0}^{\infty}\binom{-(1+\mu)-\mu M}{M}(W^2)^M} \quad , \tag{4.28}$$

where the index substitution $M=k-1$ was applied.

The previous extensive calculation can be simplified with the use of powerful relations (3.1) and (3.2). Eq. (4.19) is then

$$\frac{\partial x(v,R)}{\partial v}/h_v = \frac{R\frac{1}{\mu W}\frac{\partial W}{\partial v}\left\{\sum_{k=0}^{\infty}\binom{-\frac{1}{2}-\mu k}{k}(W^2)^k - \sum_{k=0}^{\infty}\binom{-\frac{1}{2}-\mu k}{k}\frac{-\frac{1}{2}}{-\frac{1}{2}-\mu k}(W^2)^k\right\}}{\frac{R}{\sqrt{1+\mu}}\frac{\partial W}{\partial v}\sqrt{\sum_{k=0}^{\infty}\binom{-\mu k-(\mu+2)}{k}(W^2)^k}} = \frac{\frac{1}{\mu W}\left\{\frac{p^{-\frac{1}{2}+1}}{(1+\mu)p-\mu} - p^{-\frac{1}{2}}\right\}}{\frac{1}{\sqrt{1+\mu}}\sqrt{\frac{p^{-(\mu+2)+1}}{(1+\mu)p-\mu}}} = \frac{\sqrt{1+\mu}}{\mu W}\frac{\left\{\frac{p^{\frac{1}{2}}}{(1+\mu)p-\mu}-p^{-\frac{1}{2}}\right\}}{\sqrt{\frac{p^{-(\mu+1)}}{(1+\mu)p-\mu}}} =$$

$$\frac{\sqrt{1+\mu}}{\mu W}\left\{\frac{p^{\frac{1}{2}}p^{\frac{\mu+1}{2}}}{(1+\mu)p-\mu} - p^{-\frac{1}{2}}p^{\frac{\mu+1}{2}}\right\}\sqrt{(1+\mu)p-\mu} =$$

$$\frac{\sqrt{1+\mu}}{\mu W}\left\{\frac{p^{\frac{\mu}{2}}p}{(1+\mu)p-\mu} - p^{\frac{\mu}{2}}\right\}\sqrt{(1+\mu)p-\mu} = \frac{\sqrt{1+\mu}}{\mu}\frac{p^{\frac{\mu}{2}}}{W}\left\{\frac{p}{(1+\mu)p-\mu} - 1\right\}\sqrt{(1+\mu)p-\mu} =$$

$$\frac{\sqrt{1+\mu}}{\mu}\frac{p^{\frac{\mu}{2}}}{W}\frac{p-[(1+\mu)p-\mu]}{(1+\mu)p-\mu}\sqrt{(1+\mu)p-\mu} = \frac{\sqrt{1+\mu}}{\mu}\frac{p^{\frac{\mu}{2}}}{W}\frac{-[\mu p-\mu]}{\sqrt{(1+\mu)p-\mu}} = -\sqrt{1+\mu}\frac{p^{\frac{\mu}{2}}}{W}\frac{p-1}{\sqrt{(1+\mu)p-\mu}} \quad .$$
$$\tag{4.29}$$

As – according to (3.3) –

$$-W^2 p^{-\mu} + p = 1 \quad \Rightarrow \quad p-1 = W^2 p^{-\mu} \quad , \tag{4.30}$$

then (also with the help of (3.1))

$$\frac{\partial x(v,R)}{\partial v}/h_v = -\sqrt{1+\mu}\frac{p^{\frac{\mu}{2}}}{W}\frac{W^2 p^{-\mu}}{\sqrt{(1+\mu)p-\mu}} = -W\sqrt{1+\mu}\frac{p^{-\frac{\mu}{2}}}{\sqrt{(1+\mu)p-\mu}} = -W\sqrt{1+\mu}\sqrt{\frac{p^{-(\mu+1)+1}}{\sqrt{(1+\mu)p-\mu}}} =$$
$$-W\sqrt{1+\mu}\sqrt{\sum_{k=0}^{\infty}\binom{-(1+\mu)-\mu k}{k}(W^2)^k} \quad . \tag{4.31}$$

It is the same result as (4.28), only found in a less number of steps. Nevertheless, the first approach is also usefull. Although the second one is shorter, it was already known at which result we aim (thanks to the first approach), which directed the individual steps of the algebraic simplifications.

The last expression can be tested for $\mu=0$. Then $W(\mu=0) = \frac{\sin v}{\cos v}$ and

$$\frac{\partial x(v,R)}{\partial v}/h_v = -W\sqrt{1+0}\sqrt{\sum_{M=0}^{\infty}\binom{-(1+0)-0M}{M}(W^2)^M} = -W\sqrt{\sum_{M=0}^{\infty}\binom{-1}{M}(W^2)^M} =$$
$$-W\sqrt{\sum_{M=0}^{\infty}\binom{-1}{M}(W^2)^M} = -W\sqrt{(1+W^2)^{-1}} = -\frac{\sin v}{\cos v}\sqrt{\frac{1}{1+\frac{\sin^2 v}{\cos^2 v}}} = -\frac{\sin v}{\cos v}\sqrt{\frac{\cos^2 v}{\cos^2 v+\sin^2 v}} = -\sin v \quad .$$
$$\tag{4.32}$$

Similar approach can be used also for *z* coordinate.

The *z*-component of the unit vector is proportional to (see (2.13) and (2.18))

$$\frac{\partial z(v,R)}{\partial v}/h_v = \frac{\frac{1}{\sqrt{1+\mu}}R\frac{1}{\mu}\frac{\partial W}{\partial v}\left\{(1+\mu)\sum_{k=0}^{\infty}\binom{-\frac{1+\mu}{2}-\mu k}{k}(W^2)^k - \sum_{k=0}^{\infty}\binom{-\frac{1+\mu}{2}-\mu k}{k}\frac{-\frac{1+\mu}{2}}{-\frac{1+\mu}{2}-\mu k}(W^2)^k\right\}}{\frac{R}{\sqrt{1+\mu}}\frac{\partial W}{\partial v}\sqrt{\sum_{k=0}^{\infty}\binom{-(\mu+2)-\mu k}{k}(W^2)^k}} = \frac{1}{\mu}\frac{\left\{(1+\mu)\frac{p^{-\frac{1+\mu}{2}+1}}{(1+\mu)p-\mu} - p^{-\frac{1+\mu}{2}}\right\}}{\sqrt{\frac{p^{-(\mu+2)+1}}{(1+\mu)p-\mu}}} =$$

$$\frac{1}{\mu}\frac{\left\{(1+\mu)\frac{p^{\frac{1}{2}}p^{-\frac{\mu}{2}}}{(1+\mu)p-\mu} - p^{-\frac{1}{2}}p^{-\frac{\mu}{2}}\right\}}{\sqrt{\frac{p^{-(\mu+1)}}{(1+\mu)p-\mu}}} = \frac{1}{\mu}\left\{(1+\mu)\frac{p^{\frac{1}{2}}p^{-\frac{\mu}{2}}p^{\frac{\mu+1}{2}}}{(1+\mu)p-\mu} - p^{-\frac{1}{2}}p^{-\frac{\mu}{2}}p^{\frac{\mu+1}{2}}\right\}\sqrt{(1+\mu)p-\mu} =$$

$$\frac{1}{\mu}\left\{\frac{(1+\mu)p}{(1+\mu)p-\mu} - 1\right\}\sqrt{(1+\mu)p-\mu} = \frac{1}{\mu}\frac{(1+\mu)p-[(1+\mu)p-\mu]}{(1+\mu)p-\mu}\sqrt{(1+\mu)p-\mu} = \frac{1}{\sqrt{(1+\mu)p-\mu}} = \sqrt{\frac{p^{-1+1}}{(1+\mu)p-\mu}}$$
. (4.33)

Then (with help of (3.1))

$$\frac{\partial z(v,R)}{\partial v}/h_v = \sqrt{\sum_{k=0}^{\infty}\binom{-1-\mu k}{k}(W^2)^k}\quad. \tag{4.34}$$

The last expression can be again tested for $\mu=0$. Then also $W(\mu=0) = \frac{\sin v}{\cos v}$ and

$$\frac{\partial z(v,R)}{\partial v}/h_v = \sqrt{\sum_{k=0}^{\infty}\binom{-1}{k}(W^2)^k} = \sqrt{(1+W^2)^{-1}} = \sqrt{\frac{1}{1+\frac{\sin^2 v}{\cos^2 v}}} = \cos v\quad, \tag{4.35}$$

where the binomial theorem (3.15) was used.

Generally, i.e. for any $\mu\geq 0$, it follows from (4.18) that

$$\hat{v} = \begin{pmatrix}\frac{\partial x_{3D}(v,R)}{\partial v}/h_v \\ \frac{\partial y_{3D}(v,R)}{\partial v}/h_v \\ \frac{\partial z_{3D}(v,R)}{\partial v}/h_v\end{pmatrix} = \begin{pmatrix}-\cos\lambda\ W\sqrt{1+\mu}\sqrt{\sum_{k=0}^{\infty}\binom{-(1+\mu)-\mu k}{k}(W^2)^k} \\ -\sin\lambda\ W\sqrt{1+\mu}\sqrt{\sum_{k=0}^{\infty}\binom{-(1+\mu)-\mu k}{k}(W^2)^k} \\ \sqrt{\sum_{k=0}^{\infty}\binom{-1-\mu k}{k}(W^2)^k}\end{pmatrix}\quad. \tag{4.36}$$

For the special case of spherical-like coordinates ($\mu=0$) the unit vector is (see (4.32) and (4.35))

$$\hat{v}(\mu=0) = \begin{pmatrix}-\cos\lambda\ \sin v \\ -\sin\lambda\ \sin v \\ \cos v\end{pmatrix}\quad, \tag{4.37}$$

which is correct considering that $v$ is not a polar angle but it changes in opposite direction than the usual polar angle of the spherical coordinates (i.e. $v$ increases when moving from the equator towards the pole), and the unit vector has thus opposite direction than the unit vector of the standard spherical coordinates.

Another partial confirmation, that the above derivation of the unit vector is correct, is that the unit vector length

$$\left[\left(-\cos\lambda\, W\sqrt{1+\mu}\, \sqrt{\sum_{k=0}^{\infty}\binom{-(1+\mu)-\mu k}{k}(W^2)^k}\right)^2 + \left(-\sin\lambda\, W\sqrt{1+\mu}\, \sqrt{\sum_{k=0}^{\infty}\binom{-(1+\mu)-\mu k}{k}(W^2)^k}\right)^2 + \left(\sqrt{\sum_{k=0}^{\infty}\binom{-1-\mu k}{k}(W^2)^k}\right)^2\right]^{1/2} .$$
(4.38)

has to be equal to one. The left-side square can be simplified to

$$W^2(1+\mu)\sum_{k=0}^{\infty}\binom{-(1+\mu)-\mu k}{k}(W^2)^k + \sum_{k=0}^{\infty}\binom{-1-\mu k}{k}(W^2)^k \quad .$$
(4.39)

When (3.1) is used with $r = W^2$, $b = -\mu$, and as $a = -(\mu+1)$ in the first term while it is $a = -1$ in the second term, it results in

$$W^2(1+\mu)\frac{p^{-(1+\mu)+1}}{(1+\mu)p-\mu} + \frac{p^{-1+1}}{(1+\mu)p-\mu} = W^2(1+\mu)\frac{p^{-\mu}}{(1+\mu)p-\mu} + \frac{1}{(1+\mu)p-\mu} \quad .$$
(4.40)

According to (3.3)

$$W^2 p^{-\mu} = p - 1 \quad ,$$
(4.41)

and thus further simplification follows:

$$(1+\mu)\frac{p-1}{(1+\mu)p-\mu} + \frac{1}{(1+\mu)p-\mu} = \frac{p-1+\mu p-\mu+1}{p+\mu p-\mu} = 1 \quad .$$
(4.42)

It can be seen that the length of the unit vector is equal to one.

### 4.3 Unit vector in λ direction

Then the norm in this case is

$$\sqrt{\left(\frac{\partial x_{3D}(\nu,R,\lambda)}{\partial\lambda}\right)^2 + \left(\frac{\partial y_{3D}(\nu,R,\lambda)}{\partial\lambda}\right)^2 + \left(\frac{\partial z_{3D}(\nu,R)}{\partial\lambda}\right)^2} = \sqrt{[x(\nu,R)]^2\left(\frac{\partial\cos\lambda}{\partial\lambda}\right)^2 + [x(\nu,R)]^2\left(\frac{\partial\sin\lambda}{\partial\lambda}\right)^2 + \left(\frac{\partial z(\nu,R)}{\partial\lambda}\right)^2} =$$
$$\sqrt{[x(\nu,R)]^2\sin^2\lambda + [x(\nu,R)]^2\cos^2\lambda + 0} = x(\nu,R) = h_\lambda \quad .$$
(4.43)

and thus the unit vector

$$\hat{\lambda} = \begin{pmatrix}\frac{\partial x_{3D}(\nu,R,\lambda)}{\partial\lambda}/h_\lambda \\ \frac{\partial y_{3D}(\nu,R,\lambda)}{\partial\lambda}/h_\lambda \\ \frac{\partial z_{3D}(\nu,R)}{\partial\lambda}/h_\lambda\end{pmatrix} = \begin{pmatrix}-x(\nu,R)\sin\lambda/h_\lambda \\ x(\nu,R)\cos\lambda/h_\lambda \\ 0\end{pmatrix} \quad .$$
(4.44)

As $x(\nu,R) = h_\lambda$, the result is simply

$$\hat{\lambda} = \begin{pmatrix}\frac{\partial x_{3D}(\nu,R,\lambda)}{\partial\lambda}/h_\lambda \\ \frac{\partial y_{3D}(\nu,R,\lambda)}{\partial\lambda}/h_\lambda \\ \frac{\partial z_{3D}(\nu,R)}{\partial\lambda}/h_\lambda\end{pmatrix} = \begin{pmatrix}-\sin\lambda \\ \cos\lambda \\ 0\end{pmatrix} \quad .$$
(4.45)

This result, valid for the SOS coordinates, is the same as for the spherical coordinates.

## 5 Unit vectors in SOS coordinate system for the large-*v* region

In this chapter, the unit vectors are derived in the large-v region. As the derivation is similar to the derivation for the small-v region (only the particular parameters in the relations vary), the derivation for the unit vectors $\widehat{R}$ and $\widehat{v}$ is placed into the **Appendix A**. The results of the derivation are reported here.

### *5.1 Unit vector in R direction*

The unit vector $\widehat{R}$ in the large-v region is

$$\widehat{R} = \begin{pmatrix} \cos\lambda \; \dfrac{W^{-\frac{1}{1+\mu}}}{\sqrt{1+\mu}} \sqrt{\sum_{k=0}^{\infty} \binom{-\frac{1}{1+\mu}+\frac{\mu}{1+\mu}k}{k} \left(W^{-\frac{2}{1+\mu}}\right)^{k}} \\ \sin\lambda \; \dfrac{W^{-\frac{1}{1+\mu}}}{\sqrt{1+\mu}} \sqrt{\sum_{k=0}^{\infty} \binom{-\frac{1}{1+\mu}+\frac{\mu}{1+\mu}k}{k} \left(W^{-\frac{2}{1+\mu}}\right)^{k}} \\ \sqrt{\sum_{k=0}^{\infty} \binom{-1+\frac{\mu}{1+\mu}k}{k} \left(W^{-\frac{2}{1+\mu}}\right)^{k}} \end{pmatrix} \quad . \tag{5.1}$$

For the special case when $\mu=0$ (i.e. equivalent to the spherical coordinates), the unit vector is

$$\widehat{R}(\mu=0) = \begin{pmatrix} \cos\lambda \cos\nu \\ \sin\lambda \cos\nu \\ \sin\nu \end{pmatrix} \quad , \tag{5.2}$$

i.e. the same as in the small-*v* region, as expected.

### *5.2 Unit vector in v direction*

The unit vector $\widehat{v}$ in the large-v region is (see **Appendix A**)

$$\widehat{v} = \begin{pmatrix} \dfrac{\partial x_{3D}(v,R)}{\partial v}/h_v \\ \dfrac{\partial y_{3D}(v,R)}{\partial v}/h_v \\ \dfrac{\partial z_{3D}(v,R)}{\partial v}/h_v \end{pmatrix} = \begin{pmatrix} -\cos\lambda \sqrt{\sum_{k=0}^{\infty} \binom{-1+\frac{\mu}{1+\mu}k}{k}\left(W^{-\frac{2}{1+\mu}}\right)^{k}} \\ -\sin\lambda \sqrt{\sum_{k=0}^{\infty} \binom{-1+\frac{\mu}{1+\mu}k}{k}\left(W^{-\frac{2}{1+\mu}}\right)^{k}} \\ \dfrac{W^{-\frac{1}{1+\mu}}}{\sqrt{1+\mu}} \sqrt{\sum_{k=0}^{\infty} \binom{-\frac{1}{1+\mu}+\frac{\mu}{1+\mu}k}{k}\left(W^{-\frac{2}{1+\mu}}\right)^{k}} \end{pmatrix} \quad . \tag{5.3}$$

The unit vector for the special case of spherical coordinates ($\mu=0$) is (see **Appendix A**)

$$\hat{v}(\mu = 0) = \begin{pmatrix} -\cos\lambda\,\sin v \\ -\sin\lambda\,\sin v \\ \cos v \end{pmatrix} \quad . \tag{5.4}$$

This is the same result as for the small-$v$ region, as expected for spherical-like coordinates.

In **Appendix A**, there is also a proof that the unit vector length is equal to one in the large-$v$ region as well, which partially confirms the correctness of its derivation.

## 5.3 Unit vector in λ direction

In the large-$v$ region, the unit vector in $\lambda$-direction is simply (see (2.32) – (2.34) and (2.37))

$$\hat{\lambda} = \begin{pmatrix} \frac{\partial x_{3D}(v,R,\lambda)}{\partial \lambda}/h_\lambda \\ \frac{\partial y_{3D}(v,R,\lambda)}{\partial \lambda}/h_\lambda \\ \frac{\partial z_{3D}(v,R)}{\partial \lambda}/h_\lambda \end{pmatrix} = \begin{pmatrix} -x(v,R)\sin\lambda/h_\lambda \\ x(v,R)\cos\lambda/h_\lambda \\ 0 \end{pmatrix} = \begin{pmatrix} -\sin\lambda \\ \cos\lambda \\ 0 \end{pmatrix} . \tag{5.5}$$

This result, valid for the SOS coordinates, is the same as for spherical coordinates.

## 6 Matrix expression of the transformation and inverse transformation

Similarly as for spherical-coordinate equivalent solution for $\mu=0$, for which the relation of the SOS unit vectors to the Cartesian unit vectors can be written in the form of matrix

$$\begin{pmatrix} \hat{R} \\ \hat{v} \\ \hat{\lambda} \end{pmatrix} = \begin{pmatrix} \cos v \cos\lambda & \cos v \sin\lambda & \sin v \\ -\sin v \cos\lambda & -\sin v \sin\lambda & \cos v \\ -\sin\lambda & \cos\lambda & 0 \end{pmatrix} \begin{pmatrix} \hat{x} \\ \hat{y} \\ \hat{z} \end{pmatrix} , \tag{6.1}$$

the same can be done for the SOS coordinates with $\mu>0$, as shown in the following paragraphs.

### 6.1 Unit vector transformation for SOS and Cartesian coordinates

The SOS unit vectors are (on the basis of relations (4.8), (4.36) and (4.44)) related – in the small-$v$ region – to the Cartesian unit vectors by the relation

$$\begin{pmatrix}\widehat{R}\\ \widehat{v}\\ \widehat{\lambda}\end{pmatrix} =$$

$$\begin{pmatrix} \cos\lambda \sqrt{\sum_{k=0}^{\infty}\binom{-1-\mu k}{k}(W^2)^k} & \sin\lambda \sqrt{\sum_{k=0}^{\infty}\binom{-1-\mu k}{k}(W^2)^k} & W\sqrt{1+\mu}\sqrt{\sum_{k=0}^{\infty}\binom{-(1+\mu)-\mu k}{k}(W^2)^k} \\ -\cos\lambda\ W\sqrt{1+\mu}\sqrt{\sum_{k=0}^{\infty}\binom{-(1+\mu)-\mu k}{k}(W^2)^k} & -\sin\lambda\ W\sqrt{1+\mu}\sqrt{\sum_{k=0}^{\infty}\binom{-(1+\mu)-\mu k}{k}(W^2)^k} & \sqrt{\sum_{k=0}^{\infty}\binom{-1-\mu k}{k}(W^2)^k} \\ -\sin\lambda & \cos\lambda & 0 \end{pmatrix}\begin{pmatrix}\widehat{x}\\ \widehat{y}\\ \widehat{z}\end{pmatrix}$$

. (6.2)

In the large-$v$ region, the SOS unit vectors are related (on the basis of relations (5.1), (5.3) and (5.5)) to the Cartesian unit vectors by the relation

$$\begin{pmatrix}\widehat{R}\\ \widehat{v}\\ \widehat{\lambda}\end{pmatrix} =$$

$$\begin{pmatrix} \cos\lambda \frac{W^{-\frac{1}{1+\mu}}}{\sqrt{1+\mu}}\sqrt{\sum_{k=0}^{\infty}\binom{-\frac{1}{1+\mu}+\frac{\mu}{1+\mu}k}{k}\left(W^{-\frac{2}{1+\mu}}\right)^k} & \sin\lambda \frac{W^{-\frac{1}{1+\mu}}}{\sqrt{1+\mu}}\sqrt{\sum_{k=0}^{\infty}\binom{-\frac{1}{1+\mu}+\frac{\mu}{1+\mu}k}{k}\left(W^{-\frac{2}{1+\mu}}\right)^k} & \sqrt{\sum_{k=0}^{\infty}\binom{-1+\frac{\mu}{1+\mu}k}{k}\left(W^{-\frac{2}{1+\mu}}\right)^k} \\ -\cos\lambda \sqrt{\sum_{k=0}^{\infty}\binom{-1+\frac{\mu}{1+\mu}k}{k}\left(W^{-\frac{2}{1+\mu}}\right)^k} & -\sin\lambda \sqrt{\sum_{k=0}^{\infty}\binom{-1+\frac{\mu}{1+\mu}k}{k}\left(W^{-\frac{2}{1+\mu}}\right)^k} & \frac{W^{-\frac{1}{1+\mu}}}{\sqrt{1+\mu}}\sqrt{\sum_{k=0}^{\infty}\binom{-\frac{1}{1+\mu}+\frac{\mu}{1+\mu}k}{k}\left(W^{-\frac{2}{1+\mu}}\right)^k} \\ -\sin\lambda & \cos\lambda & 0 \end{pmatrix}\begin{pmatrix}\widehat{x}\\ \widehat{y}\\ \widehat{z}\end{pmatrix}$$

. (6.3)

This extensive matrix expressions can be simplified – in the small-$v$ region – with the notation

$$f_1 = \sqrt{\sum_{k=0}^{\infty}\binom{-1-\mu k}{k}(W^2)^k} \quad \text{and} \quad f_2 = W\sqrt{1+\mu}\sqrt{\sum_{k=0}^{\infty}\binom{-(1+\mu)-\mu k}{k}(W^2)^k} \quad , \tag{6.4}$$

and – in the large-$v$ region – with the notation

$$f_1 = \frac{W^{-\frac{1}{1+\mu}}}{\sqrt{1+\mu}}\sqrt{\sum_{k=0}^{\infty}\binom{-\frac{1}{1+\mu}+\frac{\mu}{1+\mu}k}{k}\left(W^{-\frac{2}{1+\mu}}\right)^k} \quad \text{and} \quad f_2 = \sqrt{\sum_{k=0}^{\infty}\binom{-1+\frac{\mu}{1+\mu}k}{k}\left(W^{-\frac{2}{1+\mu}}\right)^k} \quad .$$

(6.5)

With the above notation, the unit vectors transformation can be written in both regions as

$$\begin{pmatrix}\widehat{R}\\ \widehat{v}\\ \widehat{\lambda}\end{pmatrix} = \begin{pmatrix} f_1\cos\lambda & f_1\sin\lambda & f_2 \\ -f_2\cos\lambda & -f_2\sin\lambda & f_1 \\ -\sin\lambda & \cos\lambda & 0 \end{pmatrix}\begin{pmatrix}\widehat{x}\\ \widehat{y}\\ \widehat{z}\end{pmatrix} \tag{6.6}$$

when taking the proper $f_1$, $f_2$ functions (see (6.4) or (6.5)) in the small- or in the large-$v$ region depending on the value of $W$ (see (2.4)). This transformation remains still analytical (although generally not in a closed form) providing that the sums in (6.4) or (6.5) are convergent. The transformation is similar to the one valid for spherical coordinates, only cosines and sines of the coordinate $v$ (parametric latitude) are exchanged by $f_1$, $f_2$ functions for SOS coordinates vector transformation.

## 6.2 Generalized sine and cosine

It does worth to note other remarkable properties of $f_1$, $f_2$ functions. It can be shown with the help of (3.1) that

$$f_1^2 + f_2^2 = 1 \quad . \tag{6.7}$$

Further, with the help of (2.3), it can be seen that

$$f_1(\nu = 0) = 1 \quad \text{and} \quad f_2(\nu = 0) = 0 \quad , \tag{6.8}$$

$$f_1\left(\nu = \frac{\pi}{2}\right) = 0 \quad \text{and} \quad f_2\left(\nu = \frac{\pi}{2}\right) = 1 \quad . \tag{6.9}$$

With the help of the binomial identities (3.12), (3.13), a relation between $f_2$ ((6.4) or (6.5)) and the $R$-coordinate metric scale factor ((2.17) or (2.35)) can be found:

$$\frac{1-h_R^2}{\mu} = \frac{f_2^2}{1+\mu} \quad . \tag{6.10}$$

Using (A6.7), further relations between $f_1$, $f_2$ and $h_R$ are derived:

$$h_R^2 = 1 - \frac{\mu}{1+\mu}f_2^2 = \frac{1}{1+\mu} + \frac{\mu}{1+\mu}f_1^2 \quad , \tag{6.11}$$

and the inverse relations

$$f_2^2 = \frac{1+\mu}{\mu}(1 - h_R^2) \quad , \qquad f_1^2 = \frac{(1+\mu)h_R^2 - 1}{\mu} \quad . \tag{6.12}$$

Specially, for $\mu \to 0$, we obtain (with the help of (2.3) and the binomial theorem)

$$\lim_{\mu \to 0}\left(\frac{1-h_R^2}{\mu}\right) = \lim_{\mu \to 0}\left(\frac{f_2^2}{1+\mu}\right) = \lim_{\mu \to 0}(f_2^2) = \sin^2\nu \quad . \tag{6.13}$$

From (A6.7) it then follows that

$$\lim_{\mu \to 0}(f_1^2) = \cos^2\nu \quad . \tag{6.14}$$

In the extreme case of spherical coordinates (i.e. when $\mu=0$), functions $f_1$, $f_2$ then became standard trigonometric functions.

Properties of derivatives of $f_1$ and $f_2$ (not shown in this article) are also found to be very similar to that ones of trigonometric functions.

The above reported properties of $f_1$ and $f_2$ lead to introduction of their notation as generalized cosine ($f_1$) and generalized sine ($f_2$) in the frame of SOS coordinate system.

With the help of (3.11) (Jensen's identity) and relations for $f_1$ and $f_2$, the following identity between metric scale factors and $f_1$, $f_2$, can be derived:

$$h_R^2 h_\nu^2 \frac{(1+\mu)}{\left(\frac{\partial W}{\partial \nu}\right)^2 R^2} = f_1^2 f_2^2 \frac{1}{W^2(1+\mu)} \quad . \tag{6.15}$$

This formula and the expressions (6.12) further help to derive the metric scale factor $h_\nu$ in terms of the powers of the metric scale factor $h_R$:

$$h_\nu^2 = \frac{\left(\frac{\partial W}{\partial \nu}\right)^2 R^2}{(1+\mu)} \frac{1}{\mu^2 W^2 h_R^2}\left(-1 + (2+\mu)h_R^2 - (1+\mu)h_R^4\right) \quad . \tag{6.16}$$

### 6.3 Inverse transformation

As both the Cartesian and the SOS coordinates are orthogonal, the matrix (6.6) is an orthogonal matrix, which means that its inverse is simply its transpose. The Cartesian unit vectors are thus related – in the small-v region – to the SOS unit vectors by the relation

$$\begin{pmatrix}\hat{x}\\\hat{y}\\\hat{z}\end{pmatrix} = \begin{pmatrix} f_1\cos\lambda & -f_2\cos\lambda & -\sin\lambda \\ f_1\sin\lambda & -f_2\sin\lambda & \cos\lambda \\ f_2 & f_1 & 0 \end{pmatrix}\begin{pmatrix}\hat{R}\\\hat{v}\\\hat{\lambda}\end{pmatrix} \quad . \tag{6.17}$$

### 6.4 Vector transformation

With the above introduced notation ($f_1$, $f_2$), the unit vectors of SOS coordinates are simplified (see (4.8), (4.36), (4.44), and (5.1), (5.3), (5.5)) to

$$\hat{R} = \begin{pmatrix} f_1\cos\lambda \\ f_1\sin\lambda \\ f_2 \end{pmatrix}, \qquad \hat{v} = \begin{pmatrix} -f_2\cos\lambda \\ -f_2\sin\lambda \\ f_1 \end{pmatrix}, \qquad \hat{\lambda} = \begin{pmatrix} -\sin\lambda \\ \cos\lambda \\ 0 \end{pmatrix} \quad . \tag{6.18}$$

Then, the vector **A**, see (1.1), for which its components $A_R$, $A_v$, $A_\lambda$ in the SOS coordinates are known, can be expressed in the Cartesian-coordinates components $A_x, A_y, A_z$ as

$$\mathbf{A} = \begin{pmatrix}A_x\\A_y\\A_z\end{pmatrix} = \begin{pmatrix} f_1\cos\lambda \\ f_1\sin\lambda \\ f_2 \end{pmatrix}A_R + \begin{pmatrix} -f_2\cos\lambda \\ -f_2\sin\lambda \\ f_1 \end{pmatrix}A_v + \begin{pmatrix} -\sin\lambda \\ \cos\lambda \\ 0 \end{pmatrix}A_\lambda =$$
$$\begin{pmatrix} A_R f_1\cos\lambda - A_v f_2\cos\lambda - A_\lambda\sin\lambda \\ A_R f_1\sin\lambda - A_v f_2\sin\lambda + A_\lambda\cos\lambda \\ A_R f_2 + A_v f_1 \end{pmatrix} \quad . \tag{6.19}$$

## 7 Examples

### 7.1.1 Special case: μ=1

One more example (along with μ=0 mentioned in the introduction to the chapter 6, and μ=2 which is not shown here) when the transformation can be written in a closed form is the case when the oblateness parameter μ=1. Then, in the small-v region,

$$f_1 = \sqrt{\sum_{k=0}^{\infty}\binom{-1-k}{k}(W^2)^k} \quad \text{and} \quad f_2 = W\sqrt{2}\sqrt{\sum_{k=0}^{\infty}\binom{-2-k}{k}(W^2)^k} \quad , \tag{7.1}$$

thus (using (3.12) with β=k+1 and therefor β+k−1=2k)

$$f_1 = \sqrt{\sum_{k=0}^{\infty}\binom{2k}{k}(-W^2)^k} \quad \text{and} \quad f_2 = W\sqrt{2}\sqrt{\sum_{k=0}^{\infty}\binom{1+2k}{k}(-W^2)^k} \quad . \tag{7.2}$$

Using further the identities (3.16) and (3.17), the result is

$$f_1 = \sqrt{\frac{1}{\sqrt{1-4(-W^2)}}} = \sqrt{\frac{1}{\sqrt{1+4W^2}}} \quad \text{and} \quad f_2 = W\sqrt{2}\sqrt{\frac{1}{\sqrt{1-4(-W^2)}}\left(\frac{1-\sqrt{1-4(-W^2)}}{2(-W^2)}\right)^1} =$$

$$W\sqrt{2}\sqrt{\frac{1}{2W^2}\frac{\sqrt{1+4W^2}-1}{\sqrt{1+4W^2}}} = \sqrt{1-\frac{1}{\sqrt{1+4W^2}}} \quad . \tag{7.3}$$

Considering that, for $\mu=1$, the parameter $W$ (see (2.3)) is equal to

$$W = \left(\frac{R}{R_0}\right)\frac{\sin v}{\cos^2 v} \quad , \tag{7.4}$$

the following relation is obtained:

$$f_1 = \sqrt{\frac{1}{\sqrt{1+4\left(\frac{R}{R_0}\right)^2\frac{\sin^2 v}{\cos^4 v}}}} \quad \text{and} \quad f_2 = \sqrt{1-\frac{1}{\sqrt{1+4\left(\frac{R}{R_0}\right)^2\frac{\sin^2 v}{\cos^4 v}}}} \quad . \tag{7.5}$$

On the border line between the small-$v$ and the large-$v$ region, eq. (2.4) is valid for $W$. Therefore, $W_{border} = \sqrt{1^1/(1+1)^{1+1}} = \frac{1}{2}$ on this line, and the value for $f_1$ and $f_2$ is there as follows: $f_1 = \sqrt{1/\sqrt{2}}$, $f_2 = \sqrt{1-1/\sqrt{2}}$. Note, that $f_1^2 + f_2^2 = 1$.

In the large-$v$ region, the relations for $\mu=1$ are

$$f_1 = \frac{1}{\sqrt{2W}}\sqrt{\sum_{k=0}^{\infty}\binom{-\frac{1}{2}+\frac{1}{2}k}{k}(W^{-1})^k} \quad \text{and} \quad f_2 = \sqrt{\sum_{k=0}^{\infty}\binom{-1+\frac{1}{2}k}{k}(W^{-1})^k} \quad . \tag{7.6}$$

For these infinite series, there is no simple well-known formula as for the series in the small-$v$ region. Therefore, the formula (3.1) has to be used to obtain a closed form. When used for the series contained in $f_1$, then

$$f_1 = \frac{1}{\sqrt{2W}}\sqrt{\sum_{k=0}^{\infty}\binom{-\frac{1}{2}+\frac{1}{2}k}{k}(W^{-1})^k} = \frac{1}{\sqrt{2W}}\sqrt{\frac{p^{-\frac{1}{2}+1}}{\left(1-\frac{1}{2}\right)p+\frac{1}{2}}} \quad \text{where} \quad -W^{-1}p^{\frac{1}{2}} + p = 1 \quad . \tag{7.7}$$

Therefore,

$$f_1 = \frac{1}{\sqrt{2W}}\sqrt{\frac{p^{\frac{1}{2}}}{\frac{1}{2}p+\frac{1}{2}}} \quad \text{where} \quad p^{\frac{1}{2}} = W(p-1) \quad , \tag{7.8}$$

and

$$f_1 = \frac{1}{\sqrt{2W}}\sqrt{2W\frac{p-1}{p+1}} = \sqrt{\frac{p-1}{p+1}} \quad \text{where} \quad W^2 p^2 - (2W^2+1)p + W^2 = 0 \quad . \tag{7.9}$$

The solution of the quadratic equation on the right side is

$$p_{1,2} = \frac{2W^2+1\pm\sqrt{4W^2+1}}{2W^2} \quad . \tag{7.10}$$

Finally,

$$f_1 = \sqrt{\frac{\frac{2W^2+1\pm\sqrt{4W^2+1}}{2W^2}-1}{\frac{2W^2+1\pm\sqrt{4W^2+1}}{2W^2}+1}} = \sqrt{\frac{1\pm\sqrt{4W^2+1}}{(4W^2+1)\pm\sqrt{4W^2+1}}} = \sqrt{\frac{1}{\sqrt{4W^2+1}}\frac{1\pm\sqrt{4W^2+1}}{\sqrt{4W^2+1}\pm 1}} = \sqrt{\frac{1}{\sqrt{4W^2+1}}}, \quad (7.11)$$

where only the positive expression under the outer radical was assumed. The relation is the same as in the small-*v* region.

When the identity (3.1) is applied in the large-*v* region for the series contained in $f_2$, the following is obtained for *μ*=1:

$$f_2 = \sqrt{\sum_{k=0}^{\infty}\binom{-1+\frac{1}{2}k}{k}(W^{-1})^k} = \sqrt{\frac{p^{-1+1}}{\left(1-\frac{1}{2}\right)p+\frac{1}{2}}} \quad \text{where} \quad W^{-1} = \frac{p-1}{p^{\frac{1}{2}}} \quad . \quad (7.12)$$

As $W^1$ is non-negative (we are restricted to the first quadrant), *p* has to be equal or larger than one. Then

$$f_2 = \sqrt{\frac{2}{p+1}} \quad \text{where} \quad p_{1,2} = \frac{2W^2+1\pm\sqrt{4W^2+1}}{2W^2} \quad . \quad (7.13)$$

As *W* in the large-*v* region is larger than ½, only the solution with + is allowed on the right side of the previous relation in order to keep *p* equal or larger than one. Therefore

$$f_2 = \sqrt{\frac{2}{\frac{2W^2+1+\sqrt{4W^2+1}}{2W^2}+1}} = \sqrt{\frac{4W^2}{(4W^2+1)+\sqrt{4W^2+1}}} = \sqrt{\frac{1}{\sqrt{4W^2+1}}\frac{(\sqrt{4W^2+1}-1)(\sqrt{4W^2+1}+1)}{\sqrt{4W^2+1}+1}} = \sqrt{1-\frac{1}{\sqrt{4W^2+1}}},$$

$$(7.14)$$

which is the same relation as in the small-*v* region.

### 7.1.2 Points at the equator and at the pole for general *μ*

The use of Eq. (6.19) can be demonstrated at a simple example. The vector **A** at the equator (i.e. in the small-*v* region) of an oblate spheroid with the equatorial radius $R_0$ (i.e. *v*=0, $R=R_0$), pointing exactly in the direction from the centre of the spheroid, and having thus components in the SOS system $(A_R, A_v, A_\lambda) = (A_R, 0, 0)$, can be written in the Cartesian coordinates using (6.19) as follows:

$$\mathbf{A} = \begin{pmatrix} A_x \\ A_y \\ A_z \end{pmatrix} = \begin{pmatrix} A_R f_1 \cos\lambda - 0 f_2 \cos\lambda - 0\sin\lambda \\ A_R f_1 \sin\lambda - 0 f_2 \sin\lambda + 0\cos\lambda \\ A_R f_2 + 0 f_1 \end{pmatrix} = \begin{pmatrix} A_R f_1 \cos\lambda \\ A_R f_1 \sin\lambda \\ A_R f_2 \end{pmatrix} \quad . \quad (7.15)$$

Assuming $R=R_0$ and *v*=0, $W=0$. Therefore (see (6.4)),

$$f_1 = \sqrt{\sum_{k=0}^{\infty}\binom{-1-\mu k}{k}(0)^k} = 1 \quad \text{and} \quad f_2 = 0\sqrt{1+\mu}\sqrt{\sum_{k=0}^{\infty}\binom{-(1+\mu)-\mu k}{k}(0)^k} = 0, \quad (7.16)$$

and, further, the SOS coordinates of the point at which the vector is acting $(R, v, \lambda) = (R_0, 0, \lambda)$ are transformed to the Cartesian coordinates (see (2.5)-(2.7)) as $(R_0\cos\lambda, R_0\sin\lambda, 0)$. Then

$$\mathbf{A}(R_0\cos\lambda, R_0\sin\lambda, 0) = \begin{pmatrix} A_x \\ A_y \\ A_z \end{pmatrix} = \begin{pmatrix} A_R \cos\lambda \\ A_R \sin\lambda \\ 0 \end{pmatrix}, \qquad (7.17)$$

as expected for the vector at the equator pointing from the centre.

Similar approach to the use of Eq. (6.19) in the large-$v$ region can be demonstrated at the pole The vector **A** at the pole of an oblate spheroid with the equatorial radius $R_0$ (i.e. $v=\frac{\pi}{2}$, $R=R_0$), pointing exactly in the direction from the centre of the spheroid (i.e. along the rotation axis), and having thus components in the SOS system again $(A_R, A_v, A_\lambda) = (A_R, 0, 0)$, can be written in the Cartesian coordinates using (6.19) again as in (7.15).

Assuming $R=R_0$ and $v=\frac{\pi}{2}$, $W^{-1} = 0$. Therefore (see (6.5)),

$$f_1 = \frac{0}{\sqrt{1+\mu}}\sqrt{\sum_{k=0}^{\infty}\binom{-\frac{1}{1+\mu}+\frac{\mu}{1+\mu}k}{k}(0)^k} = 0 \quad \text{and} \quad f_2 = \sqrt{\sum_{k=0}^{\infty}\binom{-1+\frac{\mu}{1+\mu}k}{k}(0)^k} = 1 \ . \quad (7.18)$$

Further, the SOS coordinates of the point on the pole at which the vector is acting $(R, v, \lambda) = \left(R_0, \frac{\pi}{2}, \lambda\right)$ are transformed to the Cartesian coordinates (see (2.23)-(2.25)) as $\left(0, 0, \frac{R_0}{\sqrt{1+\mu}}\right)$. Then

$$\mathbf{A}\left(0, 0, \frac{R_0}{\sqrt{1+\mu}}\right) = \begin{pmatrix} A_x \\ A_y \\ A_z \end{pmatrix} = \begin{pmatrix} 0 \\ 0 \\ A_R \end{pmatrix} \qquad , \qquad (7.19)$$

as expected for the vector at the pole pointing from the centre.

### 7.1.3   *Special vector field transformation*

As a last example, the vector field with a particular special dependence on the position in the SOS coordinates is transformed to the Cartesian coordinates. As the particular dependence,

$$\mathbf{A}(R, v, \lambda) = \begin{pmatrix} A_R \\ A_v \\ A_\lambda \end{pmatrix} = \begin{pmatrix} -C\frac{R}{h_R} \\ 0 \\ 0 \end{pmatrix} \qquad (7.20)$$

is selected, where *C* is a constant. This means that only a component of the vector field perpendicular to the similar oblate spheroid surfaces exists, and, moreover, it has a special dependence on *R* and *v*, determined by (2.17) or by (2.35), depending on the region. In Cartesian coordinates, the vector field is thus

$$\mathbf{A} = \begin{pmatrix} A_x \\ A_y \\ A_z \end{pmatrix} = \begin{pmatrix} f_1 \cos\lambda \\ f_1 \sin\lambda \\ f_2 \end{pmatrix} A_R = -C \begin{pmatrix} R\frac{f_1}{h_R}\cos\lambda \\ R\frac{f_1}{h_R}\sin\lambda \\ R\frac{f_2}{h_R} \end{pmatrix} \qquad . \qquad (7.21)$$

In the small-$v$ region, with help of (2.17),

$$\frac{f_1}{h_R} = \frac{\sqrt{\sum_{k=0}^{\infty} \binom{-1-\mu k}{k}(W^2)^k}}{\sqrt{\sum_{k=0}^{\infty} \binom{-\mu k}{k}(W^2)^k}} = \sqrt{\sum_{k=0}^{\infty} \frac{-1}{-1-\mu k}\binom{-1-\mu k}{k}(W^2)^k}, \quad (7.22)$$

and

$$\frac{f_2}{h_R} = \frac{W\sqrt{1+\mu}\sqrt{\sum_{k=0}^{\infty}\binom{-(1+\mu)-\mu k}{k}(W^2)^k}}{\sqrt{\sum_{k=0}^{\infty}\binom{-\mu k}{k}(W^2)^k}} = W\sqrt{1+\mu}\sqrt{\sum_{k=0}^{\infty}\frac{-(1+\mu)}{-(1+\mu)-\mu k}\binom{-(1+\mu)-\mu k}{k}(W^2)^k}, \quad (7.23)$$

where the relation (3.10) was used. With the further help of the identity (3.2), the square root can be removed:

$$\frac{f_1}{h_R} = \sqrt{\sum_{k=0}^{\infty}\frac{-1}{-1-\mu k}\binom{-1-\mu k}{k}(W^2)^k} = \sqrt{p^{-1}} = p^{-\frac{1}{2}} = \sum_{k=0}^{\infty}\frac{-\frac{1}{2}}{-\frac{1}{2}-\mu k}\binom{-\frac{1}{2}-\mu k}{k}(W^2)^k = \frac{x(v,R)}{R}, \quad (7.24)$$

where (2.5) was used in the last simplification, and

$$\frac{f_2}{h_R} = W\sqrt{1+\mu}\sqrt{\sum_{k=0}^{\infty}\frac{-(1+\mu)}{-(1+\mu)-\mu k}\binom{-(1+\mu)-\mu k}{k}(W^2)^k} = W\sqrt{1+\mu}\sqrt{p^{-(1+\mu)}} = W\sqrt{1+\mu}\, p^{-\frac{1+\mu}{2}} =$$
$$W\sqrt{1+\mu}\sum_{k=0}^{\infty}\frac{-\frac{1+\mu}{2}}{-\frac{1+\mu}{2}-\mu k}\binom{-\frac{1+\mu}{2}-\mu k}{k}(W^2)^k = \frac{(1+\mu)z(v,R)}{R}, \quad (7.25)$$

where (2.7) was used in the last simplification.

Similarly, in the large-$v$ region, with help of (2.34)

$$\frac{f_1}{h_R} = \frac{\frac{W^{-\frac{1}{1+\mu}}}{\sqrt{1+\mu}}\sqrt{\sum_{k=0}^{\infty}\binom{-\frac{1}{1+\mu}+\frac{\mu}{1+\mu}k}{k}\left(W^{-\frac{2}{1+\mu}}\right)^k}}{\frac{1}{\sqrt{1+\mu}}\sqrt{\sum_{k=0}^{\infty}\binom{\frac{\mu}{1+\mu}k}{k}\left(W^{-\frac{2}{1+\mu}}\right)^k}} = W^{-\frac{1}{1+\mu}}\sqrt{\sum_{k=0}^{\infty}\frac{-\frac{1}{1+\mu}}{-\frac{1}{1+\mu}+\frac{\mu}{1+\mu}k}\binom{-\frac{1}{1+\mu}+\frac{\mu}{1+\mu}k}{k}\left(W^{-\frac{2}{1+\mu}}\right)^k}, \quad (7.26)$$

and

$$\frac{f_2}{h_R} = \frac{\sqrt{\sum_{k=0}^{\infty}\binom{-1+\frac{\mu}{1+\mu}k}{k}\left(W^{-\frac{2}{1+\mu}}\right)^k}}{\frac{1}{\sqrt{1+\mu}}\sqrt{\sum_{k=0}^{\infty}\binom{\frac{\mu}{1+\mu}k}{k}\left(W^{-\frac{2}{1+\mu}}\right)^k}} = \sqrt{1+\mu}\sqrt{\sum_{k=0}^{\infty}\frac{-1}{-1+\frac{\mu}{1+\mu}k}\binom{-1+\frac{\mu}{1+\mu}k}{k}\left(W^{-\frac{2}{1+\mu}}\right)^k}. \quad (7.27)$$

With the further help of the identity (3.2), the square root can be removed also here:

$$\frac{f_1}{h_R} = W^{-\frac{1}{1+\mu}}\sqrt{\sum_{k=0}^{\infty}\frac{-\frac{1}{1+\mu}}{-\frac{1}{1+\mu}+\frac{\mu}{1+\mu}k}\binom{-\frac{1}{1+\mu}+\frac{\mu}{1+\mu}k}{k}\left(W^{-\frac{2}{1+\mu}}\right)^k} = W^{-\frac{1}{1+\mu}}\sqrt{p^{-\frac{1}{1+\mu}}} = W^{-\frac{1}{1+\mu}}p^{-\frac{1}{2}\frac{1}{1+\mu}} =$$
$$W^{-\frac{1}{1+\mu}}\sum_{k=0}^{\infty}\frac{-\frac{1}{2}\frac{1}{1+\mu}}{-\frac{1}{2}\frac{1}{1+\mu}+\frac{\mu}{1+\mu}k}\binom{-\frac{1}{2}\frac{1}{1+\mu}+\frac{\mu}{1+\mu}k}{k}\left(W^{-\frac{2}{1+\mu}}\right)^k = \frac{x(v,R)}{R}, \quad (7.28)$$

where (2.23) was used in the last simplification, and

$$\frac{f_2}{h_R} = \sqrt{1+\mu} \sqrt{\sum_{k=0}^{\infty} \frac{-1}{-1+\frac{\mu}{1+\mu}k} \binom{-1+\frac{\mu}{1+\mu}k}{k} \left(W^{-\frac{2}{1+\mu}}\right)^k} = \sqrt{1+\mu}\sqrt{p^{-1}} = \sqrt{1+\mu}\, p^{-\frac{1}{2}} = \frac{(1+\mu)z(\nu,R)}{R},$$
(7.29)

where (2.25) was used in the last simplification.

It can be seen that the relations looks the same in the small- and in the large-$\nu$ regions, only the proper expressions corresponding to the respective regions has to be used for $x$ and $z$.

Then, (7.21) can be rewritten to the form

$$\mathbf{A} = \begin{pmatrix} A_x \\ A_y \\ A_z \end{pmatrix} = -C \begin{pmatrix} x(\nu,R)\cos\lambda \\ x(\nu,R)\sin\lambda \\ (1+\mu)z(\nu,R) \end{pmatrix} .$$
(7.30)

Clearly, a linear dependence exists of the components of the special vector field **A** on the Cartesian coordinates $x$ and $z$. Moreover, the $A_x$ (and $A_y$) component depends linearly only on $x$ coordinate (and not on $z$ coordinate), while the $A_z$ component depends linearly only on $z$ coordinate (and not on $x$ coordinate). This is a notable result, as the same magnitude of the vector **A** projection to the *x-y* plane exists, i.e. $\sqrt{A_x^2 + A_y^2} = -Cx(\nu,R)$, for the same distance $\sqrt{x_{3D}^2 + y_{3D}^2} = x(\nu,R)$ from the rotation axis, regardless of the height above the equatorial plane (i.e. regardless the $z$-coordinate value). The vector **A** projection to the *x-y* plane thus possess cylindrical symmetry, whereas the vector **A** *z*-component increases linearly when moving from the equator upwards (remember, that the solution (7.30) is for the first quadrant only, but – due to the symmetry of the problem – the solution under the equatorial plane is a reflection of the solution above the equatorial plane).

## 8    Conclusions

The unit vectors in SOS coordinates were found (see (6.18)). The expressions employ power series with generalized binomial coefficients which are reported in (6.4) and (6.5). The unit vector transformation between the Cartesian and the SOS coordinates, and vice versa, was derived, see (6.6) and (6.17). The obtained formulas can help to transform vector fields between the two types of orthogonal coordinates, Cartesian and SOS (see (6.19)). It would advantageously simplify the problems for a special geometry when, for example, the iso-density levels of an object are of similar oblate spheroidal shape, and a vector field is associated with such object. As a by-product, generalized cosine and sine functions applicable in the Similar Oblate Spheroidal coordinate system are introduced. They are important e.g. for Laplace equation solution in SOS system, which will be shown elsewhere. Several examples demonstrated the use of the derived transformations, including the case where the vector field has a very special dependence on the position.

# Appendix A: Unit vectors in SOS coordinate system for the large-v region

In this Appendix, the unit vectors $\hat{R}$ and $\hat{v}$ are derived in the large-v region.

## A.1 Unit vector in R direction

First, the component containing derivative of *x* with respect to *R*, see (2.26) and (2.35), is derived:

$$\frac{\partial x(v,R)}{\partial R}/h_R = \frac{\frac{W^{-\frac{1}{1+\mu}}}{1+\mu}\sum_{k=0}^{\infty}\binom{-\frac{1}{2}\frac{1}{1+\mu}+\frac{\mu}{1+\mu}k}{k}\left(W^{-\frac{2}{1+\mu}}\right)^k}{\frac{1}{\sqrt{1+\mu}}\sqrt{\sum_{k=0}^{\infty}\binom{\frac{\mu}{1+\mu}k}{k}\left(W^{-\frac{2}{1+\mu}}\right)^k}} \quad . \tag{A.1}$$

When Cauchy product combined with Jensen identity (3.11) is used for the numerator (with $b= \mu/(1+\mu)$, $a=c= -\frac{1}{2}/(1+\mu)$, $t=\frac{1}{2}/(1+\mu)$), then

$$\sum_{k=0}^{\infty}\binom{-\frac{1}{2}\frac{1}{1+\mu}+\frac{\mu}{1+\mu}k}{k}\left(W^{-\frac{2}{1+\mu}}\right)^k =$$

$$\sqrt{\sum_{k=0}^{\infty}\binom{-\frac{1}{2}\frac{1}{1+\mu}+\frac{\mu}{1+\mu}k}{k}\left(W^{-\frac{2}{1+\mu}}\right)^k \sum_{k=0}^{\infty}\binom{-\frac{1}{2}\frac{1}{1+\mu}+\frac{\mu}{1+\mu}k}{k}\left(W^{-\frac{2}{1+\mu}}\right)^k} =$$

$$\sqrt{\sum_{k=0}^{\infty}\binom{\frac{\mu}{1+\mu}k}{k}\left(W^{-\frac{2}{1+\mu}}\right)^k \sum_{k=0}^{\infty}\binom{-\frac{1}{1+\mu}+\frac{\mu}{1+\mu}k}{k}\left(W^{-\frac{2}{1+\mu}}\right)^k} \quad , \tag{A.2}$$

and thus

$$\frac{\partial x(v,R)}{\partial R}/h_R = \frac{\frac{W^{-\frac{1}{1+\mu}}}{1+\mu}\sqrt{\sum_{k=0}^{\infty}\binom{\frac{\mu}{1+\mu}k}{k}\left(W^{-\frac{2}{1+\mu}}\right)^k \sum_{k=0}^{\infty}\binom{-\frac{1}{1+\mu}+\frac{\mu}{1+\mu}k}{k}\left(W^{-\frac{2}{1+\mu}}\right)^k}}{\frac{1}{\sqrt{1+\mu}}\sqrt{\sum_{k=0}^{\infty}\binom{\frac{\mu}{1+\mu}k}{k}\left(W^{-\frac{2}{1+\mu}}\right)^k}} =$$

$$\frac{W^{-\frac{1}{1+\mu}}}{\sqrt{1+\mu}}\sqrt{\sum_{k=0}^{\infty}\binom{-\frac{1}{1+\mu}+\frac{\mu}{1+\mu}k}{k}\left(W^{-\frac{2}{1+\mu}}\right)^k} \quad . \tag{A.3}$$

Similarly for *z* coordinate:

$$\frac{\partial z(v,R)}{\partial R}/h_R = \frac{\frac{1}{\sqrt{1+\mu}}\sum_{k=0}^{\infty}\binom{-\frac{1}{2}+\frac{\mu}{1+\mu}k}{k}\left(W^{-\frac{2}{1+\mu}}\right)^k}{\frac{1}{\sqrt{1+\mu}}\sqrt{\sum_{k=0}^{\infty}\binom{\frac{\mu}{1+\mu}k}{k}\left(W^{-\frac{2}{1+\mu}}\right)^k}} \quad . \tag{A.4}$$

When – again – Cauchy product combined with Jensen identity (3.11) is used for the numerator (with $b= \mu/(1+\mu)$, $a=c= -\frac{1}{2}$, $t=\frac{1}{2}$), the result is

$$\frac{\partial z(v,R)}{\partial R}/h_R = \frac{\frac{1}{\sqrt{1+\mu}}\sqrt{\sum_{k=0}^{\infty}\binom{\frac{\mu}{1+\mu}k}{k}\left(W^{-\frac{2}{1+\mu}}\right)^k}\sqrt{\sum_{k=0}^{\infty}\binom{-1+\frac{\mu}{1+\mu}k}{k}\left(W^{-\frac{2}{1+\mu}}\right)^k}}{\frac{1}{\sqrt{1+\mu}}\sqrt{\sum_{k=0}^{\infty}\binom{\frac{\mu}{1+\mu}k}{k}\left(W^{-\frac{2}{1+\mu}}\right)^k}} = \sqrt{\sum_{k=0}^{\infty}\binom{-1+\frac{\mu}{1+\mu}k}{k}\left(W^{-\frac{2}{1+\mu}}\right)^k} \quad .$$

(A.5)

The unit vector $\widehat{R}$ is then

$$\widehat{R} = \begin{pmatrix} \cos\lambda \, \frac{W^{-\frac{1}{1+\mu}}}{\sqrt{1+\mu}}\sqrt{\sum_{k=0}^{\infty}\binom{-\frac{1}{1+\mu}+\frac{\mu}{1+\mu}k}{k}\left(W^{-\frac{2}{1+\mu}}\right)^k} \\ \sin\lambda \, \frac{W^{-\frac{1}{1+\mu}}}{\sqrt{1+\mu}}\sqrt{\sum_{k=0}^{\infty}\binom{-\frac{1}{1+\mu}+\frac{\mu}{1+\mu}k}{k}\left(W^{-\frac{2}{1+\mu}}\right)^k} \\ \sqrt{\sum_{k=0}^{\infty}\binom{-1+\frac{\mu}{1+\mu}k}{k}\left(W^{-\frac{2}{1+\mu}}\right)^k} \end{pmatrix} \quad .$$

(A.6)

For the special case when $\mu=0$ (i.e. equivalent to the spherical coordinates), the parameter $W$, see (2.3), is simplified, $(\mu = 0) = \frac{\sin v}{\cos v}$, and it is obtained (thanks to the binomial theorem (3.15))

$$\frac{W^{-\frac{1}{1+\mu}}}{\sqrt{1+\mu}}\sqrt{\sum_{k=0}^{\infty}\binom{-\frac{1}{1+\mu}+\frac{\mu}{1+\mu}k}{k}\left(W^{-\frac{2}{1+\mu}}\right)^k} = W^{-1}\sqrt{\sum_{k=0}^{\infty}\binom{-1}{k}\left(W^{-2}\right)^k} = W^{-1}\sqrt{(1+W^{-2})^{-1}} =$$

$$W^{-1}\sqrt{\frac{1}{1+W^{-2}}} = \frac{\cos v}{\sin v}\sqrt{\frac{1}{1+\frac{\cos^2 v}{\sin^2 v}}} = \frac{\cos v}{\sin v}\sqrt{\frac{\sin^2 v}{\cos^2 v+\sin^2 v}} = \cos v \quad , \tag{A.7}$$

and similarly

$$\sqrt{\sum_{k=0}^{\infty}\binom{-1+\frac{\mu}{1+\mu}k}{k}\left(W^{-\frac{2}{1+\mu}}\right)^k} = \sqrt{\sum_{k=0}^{\infty}\binom{-1}{k}(W^{-2})^k} = \sin v \quad . \tag{A.8}$$

The unit vector for the special case of spherical-like coordinates ($\mu=0$) is thus

$$\widehat{R}(\mu=0) = \begin{pmatrix} \cos\lambda\cos v \\ \sin\lambda\cos v \\ \sin v \end{pmatrix} \quad , \tag{A.9}$$

the same as in the small-$v$ region, as expected.

### *A.2 Unit vector in v direction*

The derivation in the large-$v$ region is very similar to the one for the small-$v$ region. The binomial identities listed in the chapter 3 can be used for the derivation. The shortest way is by the use of (3.1) and (3.2) identities. The first component of the unit vector is proportional to (see (2.29) and (2.36))

$$\frac{\partial x(v,R)}{\partial v}\Big/h_v = \frac{R\frac{1}{\mu W^{1+\mu}}\frac{\partial W}{\partial v}\left\{\frac{1}{1+\mu}\Sigma_{k=0}^{\infty}\binom{-\frac{1}{2}\frac{1}{1+\mu}+\frac{\mu}{1+\mu}k}{k}\left(W^{-\frac{2}{1+\mu}}\right)^k - \Sigma_{k=0}^{\infty}\binom{-\frac{1}{2}\frac{1}{1+\mu}+\frac{\mu}{1+\mu}k}{k}\frac{-\frac{1}{2}\frac{1}{1+\mu}}{-\frac{1}{2}\frac{1}{1+\mu}+\frac{\mu}{1+\mu}k}\left(W^{-\frac{2}{1+\mu}}\right)^k\right\}}{\frac{R}{1+\mu}W^{-\frac{2+\mu}{1+\mu}}\frac{\partial W}{\partial v}\sqrt{\Sigma_{k=0}^{\infty}\binom{-\frac{2+\mu}{1+\mu}+\frac{\mu}{1+\mu}k}{k}\left(W^{-\frac{2}{1+\mu}}\right)^k}} =$$

$$\frac{1+\mu}{\mu}\frac{\frac{1}{1+\mu}\frac{p^{-\frac{1}{2}\frac{1}{1+\mu}+1}}{\left(1-\frac{\mu}{1+\mu}\right)p+\frac{\mu}{1+\mu}} - p^{-\frac{1}{2}\frac{1}{1+\mu}}}{\sqrt{\frac{p^{-\frac{2+\mu}{1+\mu}+1}}{\left(1-\frac{\mu}{1+\mu}\right)p+\frac{\mu}{1+\mu}}}} \quad . \tag{A.10}$$

where the identities (3.1) and (3.2) were used. This can be further simplified to

$$\frac{\partial x(v,R)}{\partial v}\Big/h_v = \frac{1+\mu}{\mu}\frac{\frac{1}{1+\mu}p^{-\frac{1}{2}\frac{1}{1+\mu}+1} - p^{-\frac{1}{2}\frac{1}{1+\mu}}\left[\left(1-\frac{\mu}{1+\mu}\right)p+\frac{\mu}{1+\mu}\right]}{\frac{p^{-\frac{1}{2}\frac{1}{1+\mu}}}{\sqrt{\left(1-\frac{\mu}{1+\mu}\right)p+\frac{\mu}{1+\mu}}}} = \frac{1+\mu}{\mu}\frac{\frac{1}{1+\mu}p-\left[\left(1-\frac{\mu}{1+\mu}\right)p+\frac{\mu}{1+\mu}\right]}{\sqrt{\left(1-\frac{\mu}{1+\mu}\right)p+\frac{\mu}{1+\mu}}} = \frac{1}{\mu}\frac{p-(1+\mu-\mu)p-\mu}{\sqrt{\left(1-\frac{\mu}{1+\mu}\right)p+\frac{\mu}{1+\mu}}} =$$

$$-\frac{1}{\sqrt{\left(1-\frac{\mu}{1+\mu}\right)p+\frac{\mu}{1+\mu}}} \quad , \tag{A.11}$$

and finally, according to (3.1), to

$$\frac{\partial x(v,R)}{\partial v}\Big/h_v = -\sqrt{\frac{p^{-1+1}}{\left(1-\frac{\mu}{1+\mu}\right)p+\frac{\mu}{1+\mu}}} = -\sqrt{\Sigma_{k=0}^{\infty}\binom{-1+\frac{\mu}{1+\mu}k}{k}\left(W^{-\frac{2}{1+\mu}}\right)^k} \quad . \tag{A.12}$$

Similar approach can be used also for *z* coordinate.

The *z*-component of the unit vector is proportional to (see (2.31) and (2.36))

$$\frac{\partial z(v,R)}{\partial v}\Big/h_v = \frac{\frac{1}{\sqrt{1+\mu}}R\frac{1}{\mu W}\frac{\partial W}{\partial v}\left\{\Sigma_{k=0}^{\infty}\binom{-\frac{1}{2}+\frac{\mu}{1+\mu}k}{k}\left(W^{-\frac{2}{1+\mu}}\right)^k - \Sigma_{k=0}^{\infty}\binom{-\frac{1}{2}+\frac{\mu}{1+\mu}k}{k}\frac{-\frac{1}{2}}{-\frac{1}{2}+\frac{\mu}{1+\mu}k}\left(W^{-\frac{2}{1+\mu}}\right)^k\right\}}{\frac{R}{1+\mu}W^{-\frac{2+\mu}{1+\mu}}\frac{\partial W}{\partial v}\sqrt{\Sigma_{k=0}^{\infty}\binom{-\frac{2+\mu}{1+\mu}+\frac{\mu}{1+\mu}k}{k}\left(W^{-\frac{2}{1+\mu}}\right)^k}} =$$

$$W^{\frac{1}{1+\mu}}\frac{\sqrt{1+\mu}}{\mu}\frac{\frac{p^{-\frac{1}{2}+1}}{\left(1-\frac{\mu}{1+\mu}\right)p+\frac{\mu}{1+\mu}} - p^{-\frac{1}{2}}}{\sqrt{\frac{p^{-\frac{2+\mu}{1+\mu}+1}}{\left(1-\frac{\mu}{1+\mu}\right)p+\frac{\mu}{1+\mu}}}} \quad . \tag{A.13}$$

where the identities (3.1) and (3.2) were used which fulfill the relation (3.3), i.e.

$$-W^{-\frac{2}{1+\mu}}p^{\frac{\mu}{1+\mu}} + p = 1 \ . \tag{A.14}$$

Then

$$\frac{\partial z(v,R)}{\partial v}/h_v = W^{\frac{1}{1+\mu}}\frac{\sqrt{1+\mu}}{\mu}\frac{p^{\frac{1}{2}} - p^{-\frac{1}{2}}\left[\left(1-\frac{\mu}{1+\mu}\right)p+\frac{\mu}{1+\mu}\right]}{\frac{p^{-\frac{1}{2}\frac{1}{1+\mu}}}{\sqrt{\left(1-\frac{\mu}{1+\mu}\right)p+\frac{\mu}{1+\mu}}}} = W^{\frac{1}{1+\mu}}\frac{\sqrt{1+\mu}}{\mu}\frac{p^{\frac{1}{2}\left(1+\frac{1}{1+\mu}\right)} - p^{\frac{1}{2}\left(-1+\frac{1}{1+\mu}\right)}\left[\left(1-\frac{\mu}{1+\mu}\right)p+\frac{\mu}{1+\mu}\right]}{\sqrt{\left(1-\frac{\mu}{1+\mu}\right)p+\frac{\mu}{1+\mu}}} =$$

$$W^{\frac{1}{1+\mu}}\frac{\sqrt{1+\mu}}{\mu}\frac{p^{\frac{1}{2}\left(1+\frac{1}{1+\mu}\right)} - \left(1-\frac{\mu}{1+\mu}\right)p^{\frac{1}{2}\left(2-1+\frac{1}{1+\mu}\right)} - \frac{\mu}{1+\mu}p^{\frac{1}{2}\left(-1+\frac{1}{1+\mu}\right)}}{\sqrt{\left(1-\frac{\mu}{1+\mu}\right)p+\frac{\mu}{1+\mu}}} = W^{\frac{1}{1+\mu}}\frac{\sqrt{1+\mu}}{\mu}\frac{\frac{\mu}{1+\mu}p^{\frac{1}{2}\left(1+\frac{1}{1+\mu}\right)} - \frac{\mu}{1+\mu}p^{\frac{1}{2}\left(-1+\frac{1}{1+\mu}\right)}}{\sqrt{\left(1-\frac{\mu}{1+\mu}\right)p+\frac{\mu}{1+\mu}}} =$$

$$W^{\frac{1}{1+\mu}}\frac{1}{\sqrt{1+\mu}}p^{\frac{1}{2}\left(-1+\frac{1}{1+\mu}\right)}\frac{p-1}{\sqrt{\left(1-\frac{\mu}{1+\mu}\right)p+\frac{\mu}{1+\mu}}} \quad . \tag{A.15}$$

As, according to (A.14),

$$p - 1 = W^{-\frac{2}{1+\mu}} p^{\frac{\mu}{1+\mu}} \quad , \tag{A.16}$$

then

$$\frac{\partial z(v,R)}{\partial v}/h_v = W^{\frac{1}{1+\mu}}\frac{1}{\sqrt{1+\mu}}p^{-\frac{1}{2}\frac{\mu}{1+\mu}}\frac{W^{-\frac{2}{1+\mu}}p^{\frac{\mu}{1+\mu}}}{\sqrt{\left(1-\frac{\mu}{1+\mu}\right)p+\frac{\mu}{1+\mu}}} = W^{-\frac{1}{1+\mu}}\frac{1}{\sqrt{1+\mu}}\frac{p^{\frac{1}{2}\frac{\mu}{1+\mu}}}{\sqrt{\left(1-\frac{\mu}{1+\mu}\right)p+\frac{\mu}{1+\mu}}} = \frac{W^{-\frac{1}{1+\mu}}}{\sqrt{1+\mu}}\sqrt{\frac{p^{-\frac{1}{1+\mu}+1}}{\left(1-\frac{\mu}{1+\mu}\right)p+\frac{\mu}{1+\mu}}} \quad .$$
(A.17)

According to the identity (3.1), the final result is

$$\frac{\partial z(v,R)}{\partial v}/h_v = \frac{W^{-\frac{1}{1+\mu}}}{\sqrt{1+\mu}}\sqrt{\sum_{k=0}^{\infty}\binom{-\frac{1}{1+\mu}+\frac{\mu}{1+\mu}k}{k}\left(W^{-\frac{2}{1+\mu}}\right)^k} \quad , \tag{A.18}$$

The unit vector $\hat{\boldsymbol{v}}$ is then

$$\hat{\boldsymbol{v}} = \begin{pmatrix} \frac{\partial x_{3D}(v,R)}{\partial v}/h_v \\ \frac{\partial y_{3D}(v,R)}{\partial v}/h_v \\ \frac{\partial z_{3D}(v,R)}{\partial v}/h_v \end{pmatrix} = \begin{pmatrix} -\cos\lambda\sqrt{\sum_{k=0}^{\infty}\binom{-1+\frac{\mu}{1+\mu}k}{k}\left(W^{-\frac{2}{1+\mu}}\right)^k} \\ -\sin\lambda\sqrt{\sum_{k=0}^{\infty}\binom{-1+\frac{\mu}{1+\mu}k}{k}\left(W^{-\frac{2}{1+\mu}}\right)^k} \\ \frac{W^{-\frac{1}{1+\mu}}}{\sqrt{1+\mu}}\sqrt{\sum_{k=0}^{\infty}\binom{-\frac{1}{1+\mu}+\frac{\mu}{1+\mu}k}{k}\left(W^{-\frac{2}{1+\mu}}\right)^k} \end{pmatrix} \quad . \tag{A.19}$$

The components of the unit vector can be tested for $\mu=0$, i.e. for spherical-like coordinates. Then also $W(\mu = 0) = \frac{\sin v}{\cos v}$, and

$$-\sqrt{\sum_{k=0}^{\infty}\binom{-1+0k}{k}\left(W^{-\frac{2}{1+0}}\right)^k} = -\sqrt{\sum_{k=0}^{\infty}\binom{-1}{k}\left(W^{-2}\right)^k} = -\sqrt{\left(1+W^{-2}\right)^{-1}} = -\sqrt{\frac{1}{1+\frac{\cos^2 v}{\sin^2 v}}} =$$

$$-\sqrt{\frac{\sin^2 v}{\sin^2 v + \cos^2 v}} = -\sin v \quad , \tag{A.20}$$

where the binomial theorem (3.15) was used. Also the last component of the unit vector can be then simplified:

$$\frac{W^{-\frac{1}{1+0}}}{\sqrt{1+0}}\sqrt{\sum_{k=0}^{\infty}\binom{-1+0k}{k}\left(W^{-\frac{2}{1+0}}\right)^k} = \frac{1}{W}\sqrt{\sum_{k=0}^{\infty}\binom{-1}{k}(W^{-2})^k} = \frac{1}{W}\sqrt{(1+W^{-2})^{-1}} = \frac{\cos\nu}{\sin\nu}\sqrt{\frac{1}{1+\frac{\cos^2\nu}{\sin^2\nu}}} =$$

$$\frac{\cos\nu}{\sin\nu}\sqrt{\frac{\sin^2\nu}{\sin^2\nu+\cos^2\nu}} = \cos\nu \quad , \qquad (A.21)$$

and the unit vector for the special case of spherical coordinates ($\mu=0$) is thus

$$\hat{v}(\mu=0) = \begin{pmatrix} -\cos\lambda\,\sin\nu \\ -\sin\lambda\,\sin\nu \\ \cos\nu \end{pmatrix} \quad . \qquad (A.22)$$

This is the same result as for small-$\nu$ region, as expected for spherical-like coordinates.

Another proof, that the above derivation of the unit vector is correct, is that the unit vector length has to be equal to one also in the large-$\nu$ region, and thus

$$\left[\left(-\cos\lambda\sqrt{\sum_{k=0}^{\infty}\binom{-1+\frac{\mu}{1+\mu}k}{k}\left(W^{-\frac{2}{1+\mu}}\right)^k}\right)^2 + \left(-\sin\lambda\sqrt{\sum_{k=0}^{\infty}\binom{-1+\frac{\mu}{1+\mu}k}{k}\left(W^{-\frac{2}{1+\mu}}\right)^k}\right)^2 + \left(\frac{W^{-\frac{1}{1+\mu}}}{\sqrt{1+\mu}}\sqrt{\sum_{k=0}^{\infty}\binom{-\frac{1}{1+\mu}+\frac{\mu}{1+\mu}k}{k}\left(W^{-\frac{2}{1+\mu}}\right)^k}\right)^2\right]^{1/2} = 1.$$

$$(A.23)$$

The left side square can be simplified to

$$\sum_{k=0}^{\infty}\binom{-1+\frac{\mu}{1+\mu}k}{k}\left(W^{-\frac{2}{1+\mu}}\right)^k + \frac{W^{-\frac{2}{1+\mu}}}{1+\mu}\sum_{k=0}^{\infty}\binom{-\frac{1}{1+\mu}+\frac{\mu}{1+\mu}k}{k}\left(W^{-\frac{2}{1+\mu}}\right)^k. \qquad (A.24)$$

When (3.1) is used with $r = W^{-\frac{2}{1+\mu}}$, $b = \mu/(\mu+1)$, and as $a=-1$ in the first term while it is $a=-1/(1+\mu)$ in the second term, it can be rewritten to the form

$$\frac{p^{-1+1}}{\left(1-\frac{\mu}{1+\mu}\right)p+\frac{\mu}{1+\mu}} + \frac{W^{-\frac{2}{1+\mu}}}{1+\mu}\frac{p^{-\frac{1}{1+\mu}+1}}{\left(1-\frac{\mu}{1+\mu}\right)p+\frac{\mu}{1+\mu}} = \frac{1}{\left(1-\frac{\mu}{1+\mu}\right)p+\frac{\mu}{1+\mu}} + \frac{W^{-\frac{2}{1+\mu}}}{1+\mu}\frac{p^{\frac{\mu}{1+\mu}}}{\left(1-\frac{\mu}{1+\mu}\right)p+\frac{\mu}{1+\mu}}. \qquad (A.25)$$

According to (3.3),

$$p^{\frac{\mu}{1+\mu}} = (p-1)W^{\frac{2}{1+\mu}} \quad , \qquad (A.26)$$

and thus further simplification follows:

$$\frac{1}{\left(1-\frac{\mu}{1+\mu}\right)p+\frac{\mu}{1+\mu}} + \frac{W^{-\frac{2}{1+\mu}}}{1+\mu}\frac{(p-1)W^{\frac{2}{1+\mu}}}{\left(1-\frac{\mu}{1+\mu}\right)p+\frac{\mu}{1+\mu}} = \frac{1}{\left(1-\frac{\mu}{1+\mu}\right)p+\frac{\mu}{1+\mu}} + \frac{1}{1+\mu}\frac{p-1}{\left(1-\frac{\mu}{1+\mu}\right)p+\frac{\mu}{1+\mu}} = \frac{1+\mu+p-1}{(1+\mu)\left[\left(1-\frac{\mu}{1+\mu}\right)p+\frac{\mu}{1+\mu}\right]} =$$

$$\frac{\mu+p}{((1+\mu)-\mu)p+\mu} = \frac{\mu+p}{p+\mu} = 1. \qquad (A.27)$$

Indeed, the length of the unit vector is equal to one.

*Acknowledgments.*

The author acknowledges support from the long-term conceptual development project of the Nuclear Physics Institute of the Czech Academy of Sciences RVO 61389005. The author declares he has no conflict of interests.

*Data Availability Statement.*

This paper deals with the derivation of theoretical relations of the vector transformation from the SOS coordinate system to the Cartesian coordinate system. Therefore, there is no dataset created within this work.